\documentclass[
twocolumn,
preprintnumbers,
 aps,
 floatfix,
 prd,
 nofootinbib,
 superscriptaddress,
 showpacs]{revtex4-1}

\usepackage[utf8]{inputenc}

\usepackage{hyperref}
\usepackage{graphicx}

\usepackage[english]{babel}

\usepackage{float}

\usepackage[printonlyused]{acronym}
\usepackage{subfigure}

\usepackage{latexsym}
\usepackage{bm}
\usepackage{amsmath}
\usepackage{amssymb}
\usepackage{amsthm}

\usepackage{placeins}

\usepackage{gensymb}

\usepackage{color}
\usepackage{slashed}
\usepackage{xcolor}
\usepackage{bbold}

\usepackage{bm,slashed,multirow,
	soul,mathtools,xspace,array,comment}  

\usepackage{psfrag,epsfig,graphicx,graphics}

\usepackage{epsf}
\usepackage{amsfonts}

\def\meson{  \pi  }

\def\slashchar#1{\setbox0=\hbox{$#1$}
	\dimen0=\wd0
	\setbox1=\hbox{/} \dimen1=\wd1
	\ifdim\dimen0>\dimen1
	\rlap{\hbox to \dimen0{\hfil/\hfil}}
	#1
	\else
	\rlap{\hbox to \dimen1{\hfil$#1$\hfil}}
	/
	\fi}

\newcommand{\APP}{App.~}
\newcommand{\APPs}{Apps.~}

\newcommand{\FIG}{Fig.~}

\newcommand{\SEC}{Sec.~}

\newcommand{\EQ}{Eq.~}
\newcommand{\EQs}{Eqs.~}

\def\tr{{\rm tr}}

\def\bei{\begin{itemize}}
	\def\ei{\end{itemize}}

\def\beeq{\begin{eqnarray}} 
	\def\beqa{\begin{eqnarray}}
		\def\bea{\begin{eqnarray}}
			
			\def\eea{\end{eqnarray}}
		\def\eqa{\end{eqnarray}}
	\def\eeeq{\end{eqnarray}}

\def\eqar{\end{array}}
\def\beqar{\begin{array}}

\def\beas{\begin{eqnarray*}}
	\def\beqas{\begin{eqnarray*}}
		
		\def\eqas{\end{eqnarray*}}
	\def\eeas{\end{eqnarray*}}

\def\beq{\begin{equation}} 
	\def\be{\begin{equation}}
		
		\def\ee{\end{equation}}
	\def\eq{\end{equation}}
\def\eeq{\end{equation}}

\def\beqd{\begin{displaymath}}
\def\eeqd{\end{displaymath}}
\def\eqd{\end{displaymath}}

\def\beeq{\begin{eqnarray}} \def\eeeq{\end{eqnarray}}

\newcommand{\fin}{\end{document}}

\def\fin{\end{document}}

\newcommand{\Msq}{ M_{\gamma \meson}^2 }

\DeclareMathOperator{\sgn}{sgn}

\begin{document}

	\title{Breakdown of collinear factorization in the exclusive photoproduction of a $  \pi ^{0}\gamma  $ pair with large invariant mass}
	
	\author{Saad Nabeebaccus}
	\email{saad.nabeebaccus@manchester.ac.uk}
	\affiliation{Universit\'e Paris-Saclay, CNRS/IN2P3, IJCLab, 91405 Orsay, France}
	\affiliation{Department of Physics \& Astronomy, University of Manchester, Manchester M13 9PL, United Kingdom}
	
	\author{Jakob Sch\"onleber}
	\email{jschoenle@bnl.gov}
	\affiliation{Institut f\"ur Theoretische Physik, Universit\"at Regensburg, D-93040 Regensburg, Germany}
 \affiliation{RIKEN BNL Research Center, Brookhaven National Laboratory, Upton, NY 11973, USA}
	
	\author{Lech Szymanowski}
	\email{Lech.Szymanowski@ncbj.gov.pl}
	\affiliation{National Centre for Nuclear Research (NCBJ), 02-093 Warsaw, Poland}
	
	\author{Samuel Wallon}
	\email{Samuel.Wallon@ijclab.in2p3.fr}
	\affiliation{Universit\'e Paris-Saclay, CNRS/IN2P3, IJCLab, 91405 Orsay, France}

	\begin{abstract}
	We study the exclusive photoproduction of a $  \pi ^{0}\gamma  $ pair with large invariant mass $ \Msq $, which is sensitive to the exchange of either two quarks or two gluons in the $ t $-channel. In this paper, we show that the process involving two-gluon exchanges does \textit{not} factorize in the Bjorken limit at the leading twist. This can be explicitly demonstrated by the fact that there exist diagrams, which contribute at the leading twist, for which \textit{Glauber gluons} are \textit{trapped}, due to the pinching of the contour integration of \textit{both} the plus and minus component of the Glauber gluon momentum. For the same reason, $\pi^0$-nucleon scattering to two photons also suffers from the same issue. On the other hand, we stress that there are no issues with respect to collinear factorization for the quark channels. By considering an analysis of all potential reduced diagrams of leading pinch-singular surfaces, we argue that the quark channel is safe from Glauber pinches, and therefore, a collinear factorization in that case follows through without any problems. This means that processes where gluon exchanges are forbidden, such as the exclusive photoproduction of $  \pi ^{\pm}\gamma  $ and $  \rho^{0,\,\pm} \gamma  $, are unaffected by the factorization breaking effects we point out in this paper.
 
	\end{abstract}
	
	\maketitle

	\section{Introduction}
	
	During the last decades, hard exclusive processes have been shown to be very promising in order to perform the 3D tomography of nucleons. The standard approach to analyze such processes is \textit{collinear factorisation}, which allows the computation of the scattering amplitude as the convolution of a hard part, calculable in perturbation theory, and other non-perturbative functions describing the transition between in/out hadronic states and partons. These include distribution amplitudes (DAs) and generalised parton distributions (GPDs).
 
 Various $ 2 \to 3  $ exclusive processes have been studied in order to probe \textit{generalized parton distributions} (GPDs)
\cite{ElBeiyad:2010pji,Boussarie:2016qop,Pedrak:2017cpp,Pedrak:2020mfm,Grocholski:2021man,Grocholski:2022rqj,Duplancic:2018bum,Duplancic:2022ffo,Duplancic:2023kwe}. A proof of factorization of such $ 2 \to 3  $ processes was recently derived in \cite{Qiu:2022bpq,Qiu:2022pla}. In addition to giving access to \textit{chiral-odd GPDs} at the leading twist \cite{ElBeiyad:2010pji,Boussarie:2016qop,Duplancic:2023kwe}, they give enhanced sensitivity for the extraction of the $ x $ dependence of GPDs \cite{Qiu:2023mrm}, beyond the ``moment'' type information, compared to the traditionally well-studied exclusive processes, such as deeply-virtual Compton scattering (DVCS) \cite{Muller:1994ses,Ji:1996nm,Radyushkin:1996nd,Radyushkin:1997ki,Collins:1998be,Moutarde:2018kwr} and deeply-virtual meson production (DVMP) \cite{Collins:1996fb,Mankiewicz:1997uy,Mankiewicz:1997aa,Vanderhaeghen:1998uc,Mankiewicz:1998kg,Frankfurt:1999fp,Vanderhaeghen:1999xj,Belitsky:2001nq,Ivanov:2004zv,Muller:2013jur,Duplancic:2016bge}.
	
An interesting case among $2 \to 3$ exclusive processes is the exclusive photoproduction of a $\pi^{0} \gamma$ pair with large invariant mass,\footnote{We note that, in line with the proof of Qiu and Yu \cite{Qiu:2022bpq,Qiu:2022pla}, what is imposed in practice is \textit{large relative transverse momenta} of the outgoing meson and photon with respect to the nucleons~\cite{Duplancic:2022ffo,Duplancic:2023kwe}.} which, due to the quantum numbers of the produced pair, is sensitive to both the exchange of quarks and gluons in the $t$-channel.	We computed the gluon-induced amplitude for this process at leading order and leading twist, assuming collinear factorisation. Surprisingly, we found that the amplitude diverges when the double convolution of the coefficient function (hard part) with the gluon GPD and distribution amplitude (DA) of the $  \pi ^{0} $ meson is performed, already at the LO. This is quite unexpected since the similar computation for the photoproduction of a $  \pi ^{\pm}\gamma  $ pair \cite{Duplancic:2018bum,Duplancic:2022ffo} and of a $  \rho ^{0,\,\pm}\gamma  $ pair \cite{Boussarie:2016qop,Duplancic:2023kwe}, which only have contributions from quark GPDs, is free of such singularities. The aim of this paper is to analyze the origin of the singularities which we discovered in the gluon GPD channel computation of the photoproduction of a $\pi^0 \gamma$ pair.

    More precisely, the singularity in the double convolution occurs when the dominant light-cone momentum components of both a gluon from the proton and a quark from the pion become much smaller than the hard scale $Q$. The singular points in loop momentum space are therefore the endpoint of the convolution integral involving the pion DA, and the points $x = \pm \xi$ between the DGLAP and ERBL region \cite{Diehl:2003ny}. The latter have been termed \textit{breakpoints} in \cite{Collins:1998be} or \textit{cross-over lines} in \cite{Muller:2014wxa}. At leading twist, the breakpoints have been found not to  cause any problems in well-established processes involving GPDs, such as DVCS and DVMP, because the integration contour is either not pinched in the breakpoint region, or its contribution is power suppressed. In this paper, we describe, for the first time, a situation where the breakpoints are responsible for the breakdown of collinear factorization at leading twist.

	We argue that the origin of the above-mentioned singularities is that the photoproduction of a $  \pi ^{0}\gamma  $ suffers from a Glauber pinch.\footnote{A similar phenomenon was also observed in pion dissociation to two jets in \cite{Braun:2001ih,Braun:2002wu}. The pinch there also corresponds to collinear-to-soft Glauber exchanges, see \SEC\ref{sec:reduceddias}.} This can be traced back to a region, contributing at the leading power in the hard scale, where a subgraph collinear to the incoming photon is present, which enables the trapping of a soft gluon in the Glauber region. The latter corresponds to a region of loop momentum $k$ where
 \begin{equation}
 \label{eq:glauber-defintion}
 |k^+ k^-| \ll |k_{\perp}^2|\,,
 \end{equation}
 in light-cone coordinates (see \SEC\ref{sec:bubble-diag-pinch}).
 
 It is important to note that this is in contrast to usual cases of ``endpoint-like'' divergences that do quite generically appear, mostly at subleading powers. In those cases, endpoint divergences are usually due to active partons\footnote{By ``active parton'', we mean here a collinear parton, connected to some non-perturbative object, that participates in the hard scattering.} becoming (ultra)soft, meaning that all components scale in the same way. In the present case, we find that the region where both partons (from the nucleon and the pion) are (ultra)soft should not be responsible for the aforementioned divergence, because it is power-suppressed, see \SEC\ref{sec:graph-d}, \SEC\ref{sec:graph-e} and \SEC\ref{sec:power-counting-soft-region}. On the other hand, we find that a region where an active gluon from the nucleon has Glauber scaling, while a quark from the pion is soft, contributes at leading power, see \SEC\ref{sec:pc-glauber}. We therefore conclude that the divergence should be attributed to this region.
 
  We stress that in processes where two-gluon exchanges in the $t$-channel are forbidden by symmetry \cite{Boussarie:2016qop,Duplancic:2018bum,Duplancic:2022ffo,Duplancic:2023kwe}, there are no factorisation breaking effects, and the proof of factorization of \cite{Qiu:2022bpq,Qiu:2022pla} still holds. On the other hand, a direct consequence of our work in this paper is that other related processes involving two-gluon exchanges, such as the exclusive production of a photon pair from $  \pi ^{0} $-nucleon scattering, considered in \cite{Qiu:2022bpq}, also suffer from the same Glauber pinch problem.
 
    The so-called ``Glauber pinch'' is an elusive property. There exist different approaches in the literature, which correspond to different interpretations of what a "Glauber pinch" or a "Glauber region" actually means.
    The Landau condition \cite{Landau:1959fi}, and its corollary, the Coleman-Norton theorem \cite{Coleman:1965xm}, are rigorous tools in determining IR singularities and the associated (pinched) regions of loop momentum space. However, the Landau condition in its standard form only provides one with the ``exact'' pinch-singular surfaces (PSSs), when certain particles are exactly massless and on-shell. In particular, it does not distinguish between soft and Glauber pinch, both of which correspond to a $0$-dimensional PSS where the loop momentum is exactly zero.
    
    The structure of this paper is as follows. In \SEC\ref{sec:review}, we briefly review important concepts relating to singularities of Feynman graphs, including the formulation of the Landau condition with an illustration on the simple example of a bubble graph. In \SEC\ref{sec:kinematics}, we introduce the necessary definitions and clarify the choice of frame. In \SEC\ref{sec:reduceddias}, we first give a brief review of the reduced graph technique for determining IR singular regions of Feynman diagrams of arbitrary order in $\alpha_s$. Subsequently, we provide a complete catalogue of superficially leading regions of the exclusive photoproduction of a $\pi^0 \gamma$ pair and discuss in depth their power counting after taking into account cancellations due to Ward identities. We will find that a specific reduced graph, \FIG\ref{fig:reduced-diagrams}(e), might be of leading power given that some momenta of the ``soft'' subgraph are pinched in a complicated configuration involving the Glauber scaling. In the following \SEC\ref{sec:example}, we show in great detail that this is indeed the case for the particular Feynman diagram in \FIG\ref{fig:explicit-2-loop}. We also show that, for this example, the genuinely ``soft'' regions (where all loop momentum components scale in the same way) is next-to-leading power. A necessary component of the power counting analysis is to take into account the possible cancellations in the sum over graphs due to gauge invariance. In \SEC\ref{sec:WIs}, we show through a detailed analysis in the Feynman gauge that the superficial leading power of the Glauber region is not reduced by such cancellations. In \SEC\ref{sec:soft-end-suppression}, we give a heuristic perturbative argument for the soft-end suppression of the pion distribution amplitude corresponding to the behaviour $\phi_{\pi}(z) \sim z$ as $z \rightarrow 0$. We conclude the main part of the paper in \SEC\ref{sec:conclusions}. In \APP\ref{app:calc}, we show the explicit tree-level result of the hard coefficient function in na\"ive collinear factorization and demonstrate that the corresponding double convolution integral is logarithmically divergent. Finally, \APP\ref{app:simple-1D-example} contains a simple one-dimensional example of a pinch, and \APP\ref{app:collinear-pinch-Landau} contains the Landau condition analysis of the collinear pinch in the off-forward kinematics.

    \section{Singularities of Feynman integrals: A brief review}
    \label{sec:review}

\subsection{General discussion}

    Consider a generic form of an $L$-loop Feynman integral in $d$ dimension with $n$ internal lines, given by
    \begin{align}
        I(z) = \lim_{\epsilon \rightarrow 0^+} \int_{\mathbb R^{dL}} d^{dL}\omega \, \frac{N(\omega,z)}{\prod_{j=0}^n (D_j(\omega,z) + i\epsilon)},
    \end{align}
 where $z$ is a collection of external momenta and masses and $\omega$  denotes the loop momentum combined into one vector. The product in the denominator runs over all internal lines of the graph, with $D_j$ being the denominator of the propagator associated to the line $j$. The functions $D_j$ and $N$ are analytic functions of the components of $\omega$ and $z$ with real coefficients. For our purposes, it is sufficient to assume that there are only massless particles with quadratic denominators, i.e.
 \begin{align}
     D_j(\omega, z) = q_j^2(\omega, z),
 \end{align}
 where $q_j$ is the four-momentum of the line $j$.
 
 Assuming that UV divergences are regularized, the integral is manifestly finite \textit{before} the limit $\epsilon \rightarrow 0^+$ is taken, since $D_j$ is real on the initial integration hyper-contour $\mathbb R^{dL}$. As $\epsilon \rightarrow 0^+$, possible PSSs $D_j = 0$ approach the integration contour, but by virtue of the multi-dimensional version of Cauchy's theorem, we might still be able to deform the initial integration hyper-contour to complex values to avoid the PSSs. Whenever one cannot make such a deformation at some point $\omega_S$, we say there is a ``pinch'' at $\omega_S$ (assuming there are no cancellations at the pinch from the numerator). 

 The Landau condition provides a rigorous way to identify pinches. The statement is: Given $z,\,\omega_S \in \mathbb R^{dL}$ such that the set
 \begin{align}
     \mathcal D = \{ j \in \{ 1,..., n \} ~|~ D_j(\omega_S, z ) = 0 \}\,,
 \end{align}
 is non-empty, we have a pinch at $\omega_S$ if and only if there exist real and non-negative numbers $\alpha_j$ for $j \in \mathcal D$ such that 
 \begin{itemize}
     \item At least one of the $\alpha_j$ is non-zero.
     \item $\forall i \in \{1,...,dL\}  ~:~ \sum_{j \in \mathcal D} \alpha_j \frac{\partial D_j}{\partial \omega_i}(\omega_S; z) = 0$. 
 \end{itemize}
 It is important to keep in mind that the existence of a pinch does \textit{not} necessarily imply the existence of a singularity of $I(z)$. It merely identifies obstacles of contour deformations, by considering only the denominators. In order to determine whether such a pinched point $\omega_S$ actually leads to a singularity, one must perform a power counting analysis which is illustrated in the following subsection.

\subsection{Bubble diagram as an illustration of pinch singularities}
\label{sec:bubble-diag-pinch}

\begin{figure}
\centering
\includegraphics[width=6.2cm]{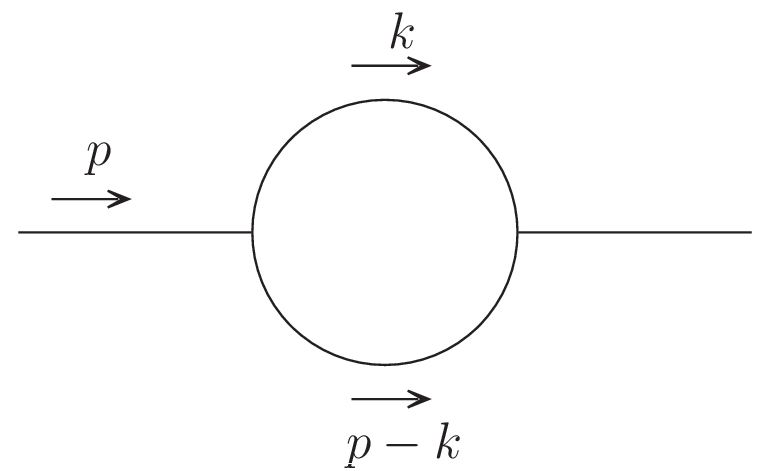}
\caption{Bubble diagram used to illustrate potential pinch singularities for different external momentum $p$.} 
\label{fig:bubble-diagram}
	\end{figure}

Consider the bubble diagram, shown in \FIG\ref{fig:bubble-diagram}, as a simple example,
 \begin{align}
 \label{eq:bubble}
   I_1(p^2) = \lim_{\epsilon\rightarrow 0^+} \int d^4k \, \frac{1}{(k^2 + i\epsilon) ((p-k)^2 + i\epsilon)}. 
 \end{align}
 First, consider the case where $p^2 \neq 0$ and $k = 0$, which means that the denominator $k^2 +i \epsilon$, as well as its first derivative, are zero. According to the Landau condition, this gives a pinch, since trivially
 \begin{align}
     \alpha_1 k = 0\,,
 \end{align}
 is true for any $\alpha_1 > 0$. However, as can be easily verified by explicit calculation, $I_1(p^2)$ is finite for $p^2 > 0$. In fact, we have here an example of a case where a pinch does not give a singularity. This can be understood through power counting. Generally speaking, a pinch implies that we have to consider a neighborhood of the pinch as a region of the loop integral, where the loop momentum is forced to have a certain asymptotic scaling depending on the PSS. By power-counting, we can estimate the size of the contribution from this subset of the loop integration region (including the loop momentum space volume). 
 
 Indeed, near $k = 0$, we can assign to $k$ a scaling for each component $k^{\mu} \sim m$,\footnote{By $\sim$, we mean an extended version of the usual asymptotic equivalence, in the sense that $f(x) \sim g(x)$ as $x \rightarrow 0$ if $f(x)/g(x) \rightarrow \text{const.}$ as $x \rightarrow 0$. The limit $x \rightarrow 0$ is omitted from notation and usually clear from the context.} where $m$ can be viewed as an IR cutoff to the integral. Then,
  \begin{align}
    \frac{1}{p^2} 
    \int_{k \sim m} d^4k \, \frac{1}{k^2} \sim \frac{1}{p^2}  \int_{k \sim m} dk \, k.
 \end{align} 
 The phase space volume gives a factor of $m^4$ and the denominator gives a factor of $\frac{1}{m^2}$ giving an overall scaling $m^2$, which is clearly finite as $m \rightarrow 0$.
 
 Consider now the case where both denominators of $I_1$ are zero, and apply the Landau condition. This gives the three equations 
\begin{eqnarray}
    && k^2 = 0, \qquad p^2 - 2 p \cdot k = 0, \qquad \alpha_1 k + \alpha_2 (k-p)=0\,, \nonumber
    \\ 
    &&\alpha_1, \alpha_2 \geq 0, \quad \alpha_1 + \alpha_2 >0\,,
 \end{eqnarray} 
 i.e.
 \begin{align}
    k^2 = 0, \qquad p^2 - 2 p \cdot k = 0, \qquad k = \alpha p, 
 \end{align}
 where $1 \geq \alpha \geq 0$. This can only have a solution if $p^2 = 0$. This is of course nothing but the well-known collinear singularity.

For later convenience, we re-analyze the integral $I_1$ using light-cone coordinates. They are defined with respect to two light-like vectors $n, \bar n$ with $n^2 = \bar n^2 = 0$ with the normalization condition $n \cdot \bar n = 1$. Without loss of generality, we choose
\begin{align}
n^{\mu} = \frac{1}{\sqrt{2}} (1,0,0,-1)^{\mu},  \qquad \bar n^{\mu} = \frac{1}{\sqrt{2}} (1,0,0,1)^{\mu}.
\end{align}
For a generic vector $V$, we define
\begin{align}
    V^+ &= n \cdot V,\nonumber\\
    V^- &= \bar n \cdot V,\nonumber\\
    V_{\perp}^\mu &= V^\mu - V^+ \bar n^\mu - V^- n^\mu.
\end{align}
We will commonly denote a vector $V$ in terms of its light-cone components by
\begin{align}
V = (V^+, V^-, V_{\perp}).
\end{align}
 It is instructive to investigate the behaviour of $I_1$ near $p^2 = 0$. For this, we go to a frame where $p$ is highly boosted in the $n$ direction, so that we have
 \begin{align}
     p \sim (\lambda^2,1, \lambda) p^-,
 \end{align}
where we investigate the asymptotic limit $\lambda \rightarrow 0$ with $p^-$ fixed. Consider the region of the loop integral where $k_{\perp} \sim \lambda p^-$. Then
\begin{align}
\label{eq:bubblepole1}
    k^2  + i\epsilon &= 2k^+ k^- + O(\lambda^2(p^-)^2)+ i\epsilon,
    \\
    \label{eq:bubblepole2}
    (p-k)^2 + i\epsilon &= - 2(p^- - k^-) k^+ + O(\lambda^2(p^-)^2) + i\epsilon.
\end{align}
Suppose that $0 < k^- < p^-$, then the poles in $k^+$ are on opposite sides of the $k^+$ contour.
Further suppose that we are in a region such that $k^- \sim p^-$ and $ p^- - k^- \sim p^-$, so that the poles in $k^+$ are separated by a distance of order $\lambda^2 p^-$. In this situation, we say that $k^+$ is \textit{trapped} to be of ${\cal O}(\lambda^2)$. We illustrate this point by a simple 1D example in \APP\ref{app:simple-1D-example}.

We can estimate the contribution from this region by using the scaling $k \sim ( \lambda^2, 1,\lambda) p^-$. The phase space volume is now given by $\lambda^4 (p^-)^4$ while the denominators in this region give $\frac{1}{\lambda^4 (p^-)^4}$, so that we get an overall estimation $\lambda^0$, corresponding to a logarithmic divergence as $\lambda \rightarrow 0$, from this region. This is of course consistent with the usual logarithmic collinear singularity of the bubble integral.

	\section{Kinematics and frame choice}
	\label{sec:kinematics}
	
	We focus here on the specific case of $  \pi ^{0}\gamma  $ photoproduction:
	\begin{align}
	\gamma (q) + N(p_{N})	 \to \gamma (q') +  \pi ^{0}(p_{\pi}) + N'(p_{N'})\;,
	\label{eq:process-def}
\end{align}
with
\begin{align}
q^2 = q'^2 = 0, \qquad p_N^2 = p_{N'}^2 = m_N^2, \qquad p_{\pi}^2 = m_{\pi}^2,
\end{align}
where $m_N$ is the nucleon mass and $m_{\pi}$ is the pion mass. We introduce the conventional notation associated to the off-forward nucleon system
\begin{align}
P = \frac{p_N + p_{N'}}{2}, \qquad \Delta = p_{N'} - p_N, \qquad t = \Delta^2. 
\end{align}
The kinematic restriction where factorization of the process in \EQ\eqref{eq:process-def} is expected to hold, according to \cite{Qiu:2022pla}, is that in the center-of-mass (CM) frame with respect to the momenta $\Delta$ and $q$, the transverse components of $q'$ and $p_{\pi}$ are much larger than $\sqrt{|t|}, m_{\pi}, m_{N}$. Consider the generic large scale to be $Q \sim \sqrt{|q_{\perp}'^2|},\sqrt{ |p_{\pi,\perp}^2| }$ in the CM frame w.r.t. $\Delta$ and $q$, and the generic small scales to be $ \sqrt{|t|}, m_{\pi}, m_N, \Lambda_{\mathrm{QCD}}$. Thus, we investigate the limit where $Q \rightarrow \infty$ with $\sqrt{|t|}, m_{\pi}, m_N, \Lambda_{\mathrm{QCD}}$ fixed. We parameterize this limit by a generic dimensionless parameter
\begin{align}
    \lambda \sim \frac{\{ \sqrt{|t|}, m_{\pi}, m_N, \Lambda_{\mathrm{QCD}} \}}{Q} \rightarrow 0.
\end{align}
For our purposes, it will be more convenient, especially for the power counting in \SEC\ref{sec:example}, to consider the CM frame with respect to $\Delta$ and $p_{\pi}$. Our choice of frame is defined by $\Delta_{\perp} = p_{\pi,\perp} = 0$. The condition of small $t$ implies that $p_N, p_{N'}$ can be viewed as approximately collinear in the $\bar n$ direction, so that we have the order of magnitude estimations
\begin{align}
\label{eq:external-scalings-nucleon}
p_N, p_{N'}, \Delta, P &\sim (1, \lambda^2, \lambda)Q, \\
p_{\pi} &\sim (\lambda^2, 1, \lambda)Q.
\label{eq:external-scalings-pion}
\end{align}
\EQ\eqref{eq:external-scalings-nucleon} implicitly assumes that the skewness
\begin{align}
    \xi = - \frac{\Delta^+}{2P^+}\,,
\end{align}
is neither very small nor very close to $1$.

On the other hand, all the momentum components of the two photons are of order $Q$ in this frame, $q, q' \sim (1,1,1) Q$, with the condition that they are real, i.e. $q^2 = q'^2 = 0$.\footnote{The photons can also be quasi-real, with $q^2 \sim q'^2 \sim \lambda^2$. This does not spoil the arguments in this paper.} Therefore, one should keep in mind that one can have singularities when virtual particles become collinear to $q$ or $q'$. Note that since the photons are physical particles, their energies, which are the sum of the plus- and minus-momenta, are positive. Together with the on-shell condition, this implies that $q^+, q^-, q'^+ ,q'^- > 0$.

\section{Scalings}
\label{sec:reduceddias}

 We start by making an analysis of the reduced diagrams that contribute to the amplitude in powers of the hard scale $Q$ \cite{Libby:1978qf,Sterman:1978bi,Sterman:1978bj,Collins:2011zzd}. Reduced diagrams are a convenient way of finding the different regions of the loop momentum integration that give leading contributions to the amplitude. These correspond to the PSSs, where some components of loop momenta are ``trapped'' by poles of denominators of internal lines, in the same spirit as discussed in \SEC\ref{sec:review}. A convenient way to identify pinches is provided by the \textit{Coleman-Norton theorem} \cite{Coleman:1965xm}, which states that a pinched configuration in loop momentum space, i.e. a PSS, corresponds to a classical scattering process. This means that the pinched (or on-shell) propagators correspond to lines which represent classical particles propagating forward with time, whereas the hard propagators are shrunk to points. In a neighborhood of such a PSS, the integrand can become very large so that, although the volume of the phase space region may be small, the region gives a large contribution to the whole integral.

Exact PSSs can be determined by the Landau conditions. Suppose an external particle (e.g. meson) has exactly a four-momentum squared of zero. Then, the Landau conditions predict an exact pinch for the loop momentum of the corresponding quark or gluon line when it is collinear, and when it is soft. In practice, external particles have non-zero four-momentum squared, for instance of order $\lambda^2$. This implies that the pinch now becomes \textit{approximate}, since the poles are now separated by a distance of $O(\lambda)$.\footnote{For example, one could define the distance as $|\delta^+| + |\delta^-| + |\delta_{\perp}|$, where $\delta$ is the difference between two pole locations and $| \cdot |$ denote the absolute value.} Still, the approximate PSSs still limit possible contour deformations, restricting loop momentum components to have certain maximum sizes.

However, note that when massless internal particles are involved, the soft pinch is always exact, independently of the momenta of external particles, and appears when the momentum of a massless propagator is zero. This is because the Landau condition gives a pinch whenever the momentum of a single massless propagator vanishes, as illustrated in \SEC\ref{sec:bubble-diag-pinch}. This might not always lead to an IR singularity because of the numerator and momentum space volume factors that may suppress this region. Whether there is indeed an IR singularity requires an analysis on a case-by-case basis. This is because there can be other denominators that become small, that compensate the suppression coming from the numerator and phase space measure. Since a generic loop momentum $k$ is always pinched at $k = 0$, one should always consider the scaling \begin{align}
k \sim (\lambda_s, \lambda_s, \lambda_s) Q,
\label{eq:soft-scaling}
\end{align}
where $\lambda_s \ll 1$. In principle, one has to consider all possible scalings of $\lambda_s$ relative to $\lambda$. Since all that matters is whether we can neglect some components of a soft momentum with respect to, say,  a collinear momentum, the only relevant cases are $\lambda_s \sim \lambda$, called soft, and $\lambda_s \sim \lambda^2$, called ultrasoft (usoft). Whether one may ``choose'' one or the other is a very subtle question. When internal particles are massless and there are only collinear or hard external particles, typically the ultrasoft region is the relevant one for fixed order Feynman graphs. This is because the loop integrals expanded in the soft region are typically scaleless and are therefore not needed to reproduce the asymptotic expansion of the loop integral in the well-known method of regions \cite{Beneke:1997zp, Jantzen:2011nz}. However, this need not be true in general, as we will see in \SEC\ref{sec:example}.

Moreover, it can be argued that in exclusive processes, the ultrasoft momentum modes are unphysical, by the virtue of confinement, in the sense that their wavelengths $\sim \frac{1}{\lambda^2Q}$ are much larger than the sizes of the hadrons involved, which are $\sim \frac{1}{\lambda Q}$. Hence, taking into account non-perturbative effects, one should consider massless propagators to be effectively cut off at virtualities around $\lambda^2 Q^2 \sim \Lambda_{\rm QCD}^2$. However, for the sake of factorization proofs, the question of whether it matters if one considers only the soft region, the ultrasoft region or both remains unclear. In fact, in most of the literature on factorization of processes involving GPDs, e.g. \cite{Collins:1996fb, Collins:1998be, Qiu:2022pla}, only the ultrasoft scaling is used to treat the ``soft'' region. This is essentially due to convenience, because it greatly simplifies the graphical power-counting, by the virtue of ultrasoft lines attaching to collinear lines without putting them off-shell. Therefore, the following all-order power-counting analysis in this section is done in the spirit of the above-mentioned works in order to exploit the same convenience.

 It is worth to mention at this point that the scaling in \EQ\eqref{eq:soft-scaling} is opposed to the \textit{Glauber scaling}, where all components are much smaller than $Q$, but the transverse momentum is larger than the product of plus and minus momenta, see \EQ\eqref{eq:glauber-defintion}. It is precisely this Glauber scaling that plays an essential role in our current paper, and leads to the breakdown of collinear factorisation in the process of exclusive $\pi^0 \gamma$ photoproduction.

In summary, the different relevant scalings for a generic momentum $k$ are\footnote{The hard-collinear scaling shown here simply corresponds to the sum of a collinear and soft momentum. We note that the meaning of hard-collinear is context dependent. To get the correct method of regions expansion, one has to take $(\lambda,1,\sqrt{\lambda})Q$ as the hard $n$-collinear scaling (and similarly for $n \leftrightarrow \bar n$).}
\begin{align}
    k &\sim Q (1, \lambda^2, \lambda) \qquad &&\bar{n}\textrm{-coll.,}\\
    k &\sim Q ( \lambda^2,1, \lambda) \qquad &&{n}\textrm{-coll.,}\\
    k &\sim Q (\lambda^2, \lambda^2, \lambda^2) \qquad &&\textrm{ultrasoft,}\\
    k &\sim Q (\lambda, \lambda, \lambda) \qquad &&\textrm{soft,}\\
    k &\sim Q (1, \lambda ,\lambda) \qquad && \textrm{hard }\bar{n}\textrm{-coll.,}\\
    k &\sim Q (\lambda, 1 ,\lambda) \qquad && \textrm{hard }n\textrm{-coll.,}\\
    \label{eq:coll-to-coll-Glauber}
    k &\sim Q (\lambda^2, \lambda^2, \lambda) \qquad &&\bar{n}\textrm{-coll. to }n\textrm{-coll. Glauber,}\\
    \label{eq:nbar-coll-to-soft-Glauber}
    k &\sim Q (\lambda,\lambda^2, \lambda) \qquad &&\bar{n}\textrm{-coll. to soft Glauber,}\\
    k &\sim Q (\lambda^2, \lambda, \lambda) \qquad &&{n}\textrm{-coll. to soft Glauber.}
\end{align}

Each Feynman diagram may have multiple regions, or equivalently reduced graphs, where we assign to each line a scaling -- hard, collinear or usoft (soft is not considered at this point) -- and correspondingly group lines and vertices together into hard, collinear or usoft subgraphs. Assuming that all lines in a subgraph have the same scaling, we can derive, by dimensional analysis and Lorentz transformation properties of Feynman graphs, an order of magnitude estimate for a given region $R$ associated to some reduced diagram. In QCD (and in Feynman gauge), the Libby-Sterman formula,  which applies only to momentum configurations with collinear, hard and usoft scalings, reads \cite{Libby:1978qf,Sterman:1978bi,Sterman:1978bj,Collins:2011zzd}
\begin{align}
\text{Contribution from }R \sim Q^{\rm p} \lambda^{\alpha},
\label{eq:pc-formula}
\end{align}
where
\begin{align}
\label{eq:pc-formula-2}
{\rm p} &= 4 - \#(\text{total number of ext. lines})\,, \notag
\\[5pt]
\alpha &= \#(\text{lines from coll. to hard}) \notag
\\
&\quad - \#(\text{scalar pol. gluons from coll. to hard}) \notag
\\
&\quad - \#(\text{ext. lines of coll.}) \notag
\\
&\quad + 2 \#(\text{gluons from usoft to hard})
\\
&\quad + 3 \#(\text{quarks from usoft to hard}) \notag
\\
&\quad + \#(\text{quarks from usoft to coll.}) \notag
\\
&\quad + \#(\text{gluons from usoft to coll.})  \notag
\\
&\quad - \#(\text{scalar pol. gluons from usoft to coll.})\,. \notag
\end{align}
Since the power of the hard scale is always equal to the mass dimension of the corresponding contribution to the amplitude, we will set $Q = 1$ throughout for notational simplicity.

We refer to gluons of ``scalar polarization'' as those associated with the leading component of the metric tensor in the propagator of the gluon connecting between subgraphs. To make this more precise, consider a gluon connecting an $\bar n$-collinear subgraph $A^{\mu}$ to a hard subgraph $H^{\mu}$, where $\mu$ is the index that is contracted by the gluon propagator of collinear momentum $l \sim (1, \lambda^2, \lambda)$. Then, by the Lorentz transformation properties of Feynman graphs, we have $A^{\mu} \sim (1, \lambda^2, \lambda)$, which can be understood by considering $A^{\mu}$ in the rest frame (where all components scale in the same way), and then boosting to the frame where $l$ is collinear. On the other hand, typically, $H^{\mu} \sim (1, 1, 1)$, since it is composed of only hard momenta.\footnote{This is discussed in Chapter 5 of \cite{Collins:2011zzd}.}

We neglect overall factors of $\lambda^{\alpha}$, since we only consider the relative scaling with respect to the $\mu$ index here. The dominant contribution in the contraction of $A^\mu$ with $H^\mu$ thus corresponds to
\begin{align}
\label{eq:dominant-AH}
   A_{\mu} H^{\mu} = A_{\mu} (n^{\mu}\bar{n}^{\nu} ) H_{\nu} + \textrm{non-dominant terms}\,.
\end{align}
We now apply the Grammer-Yennie decomposition,
\begin{align}
\label{eq:GY-AH}
A_{\mu} H^{\mu} = A_{\mu} (K^{\mu \nu} + G^{\mu \nu} ) H_{\nu}\,,
\end{align}
where the $K^{\mu\nu}$ term extracts the dominant contribution. 
From \EQ\eqref{eq:dominant-AH}, and in preparation for later use in Ward identities, we write\footnote{The $l^+$ in the denominator needs to be equipped a $i\epsilon$ prescription that is compatible with the contour deformation out of the Glauber region, if possible. This prescription is tacitly implied throughout this text.}
\begin{align}
\label{eq:GY-decomp-1}
    K^{\mu \nu} = \frac{n^{\mu} l^{\nu}}{l^+}\,,\qquad G^{\mu \nu} = g^{\mu \nu} - K^{\mu \nu}\,.
\end{align}
For an usoft $S$-to-$A$ gluon of momentum $l \sim (\lambda^2, \lambda^2, \lambda^2)$, where $S^{\mu} \sim (1,1,1)$ (up to an overall power of $\lambda$) is a usoft subgraph, we correspondingly decompose
\begin{align}
A_{\mu} S^{\mu} = A_{\mu} (\bar K^{\mu \nu} + \bar G^{\mu \nu} ) S_{\nu},
\label{eq:GY-AS}
\end{align}
where
\begin{align}
\bar K^{\mu \nu} = \frac{l^{\mu} \bar n^{\nu}}{l^-}, \qquad \bar G^{\mu \nu} = g^{\mu \nu} - \bar K^{\mu \nu}.
\label{eq:GY-decomp-2}
\end{align}
Of course, we could also have chosen $K^{\mu\nu} = \frac{n^{\mu}l^\nu}{l^+}$. However, the specific structure of $\bar{K}^{\mu\nu}$ leads to Ward identity cancellations after summing over all possible insertions of the gluon to the subgraph $A^\mu$. This is in contrast to the $A$-to-$H$ case in \EQs\eqref{eq:dominant-AH} to \eqref{eq:GY-decomp-1}, where the corresponding Ward identity cancellations occur as a result of summing over all possible gluon attachments to the $H$ subgraph. The specific choice for the structure of tensor $K$ is also linked to how we group subgraphs together, see \SEC\ref{sec:graph-c}.

Similar decompositions apply for collinear subgraphs in a direction other than $\bar n$. Thus, for $B$-to-$H$ and $S$-to-$B$ gluons, we have the same \EQs\eqref{eq:GY-decomp-1} and \eqref{eq:GY-decomp-2} with $n \leftrightarrow \bar n$ and $l^+ \leftrightarrow l^-$, where $B$ is an $ n$-collinear subgraph.

By construction, the $G (\bar{G})$ term is suppressed relative to the $K (\bar{K})$ term here. We will therefore refer to the scalar polarized gluons ($K$ or $\bar K$ terms) as ``$K$-gluons'' and to the non-scalar polarized gluons ($G$ or $\bar G$ terms)  as ``$G$-gluons''.  We stress that the  relative suppression of $G$-gluons with respect to $K$-gluons is not true in general if they are trapped in the Glauber region, as discussed in \SEC\ref{sec:WIs}.

	\begin{figure}
\centering
\includegraphics[width=4.2cm]{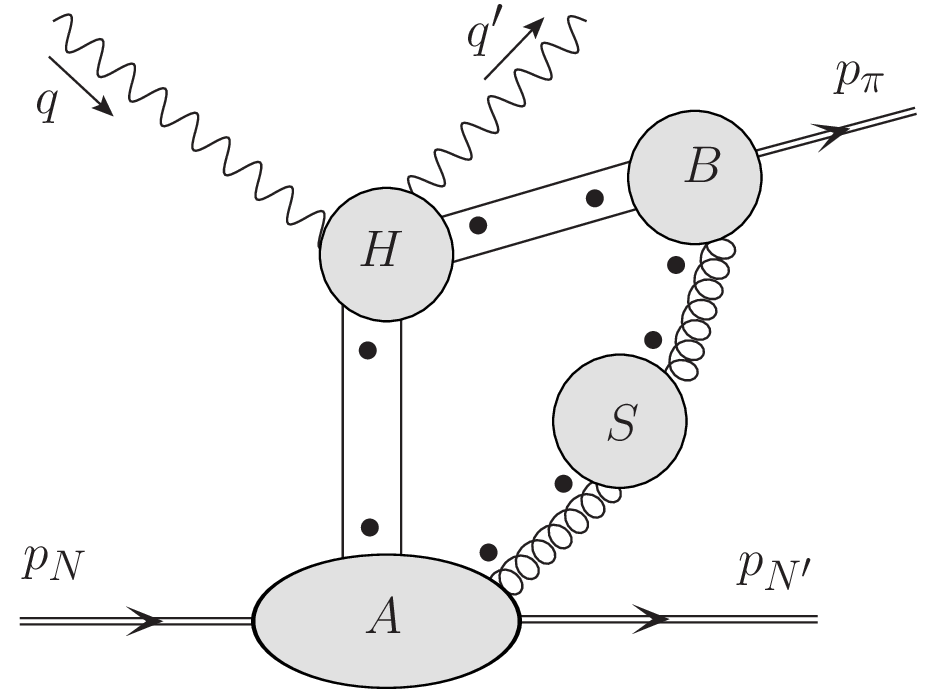}
\includegraphics[width=4.2cm]{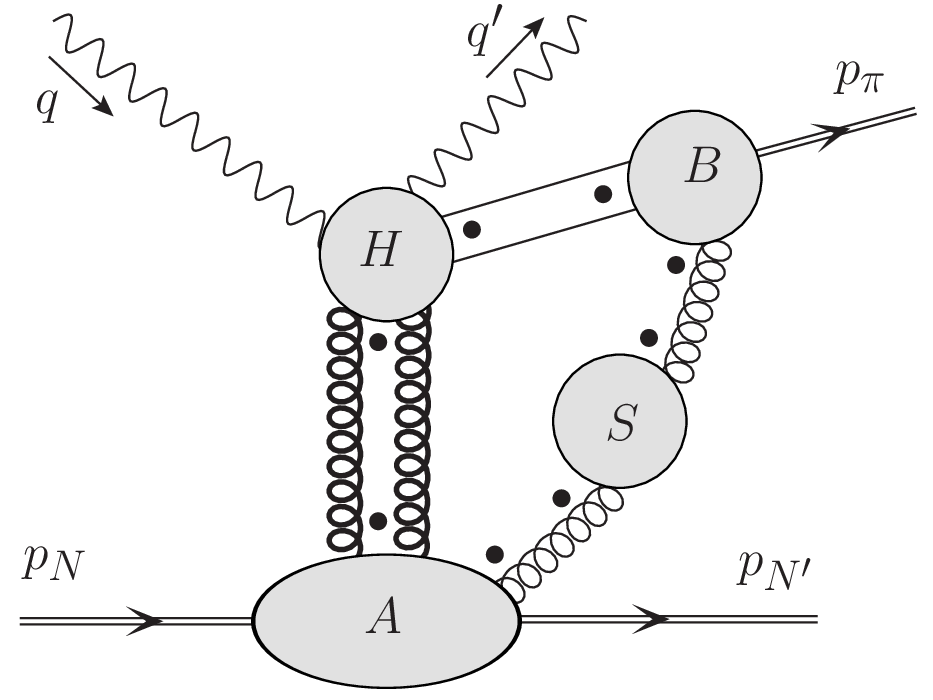}
(a)\hspace{4cm}(b)\hspace{0.4cm}
\vskip 5mm
\includegraphics[width=4.2cm]{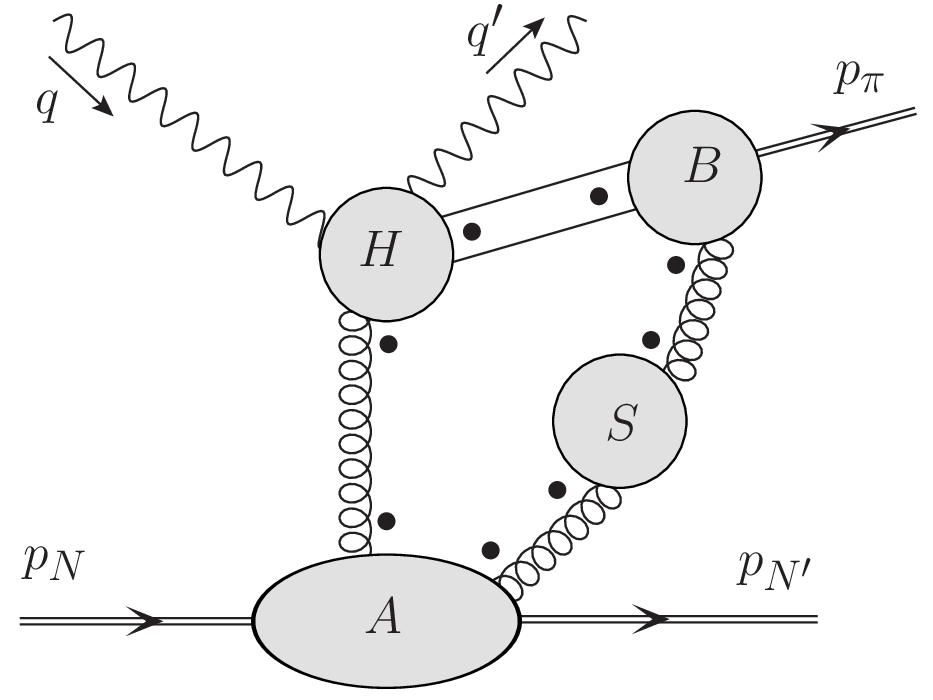}
\includegraphics[width=4.2cm]{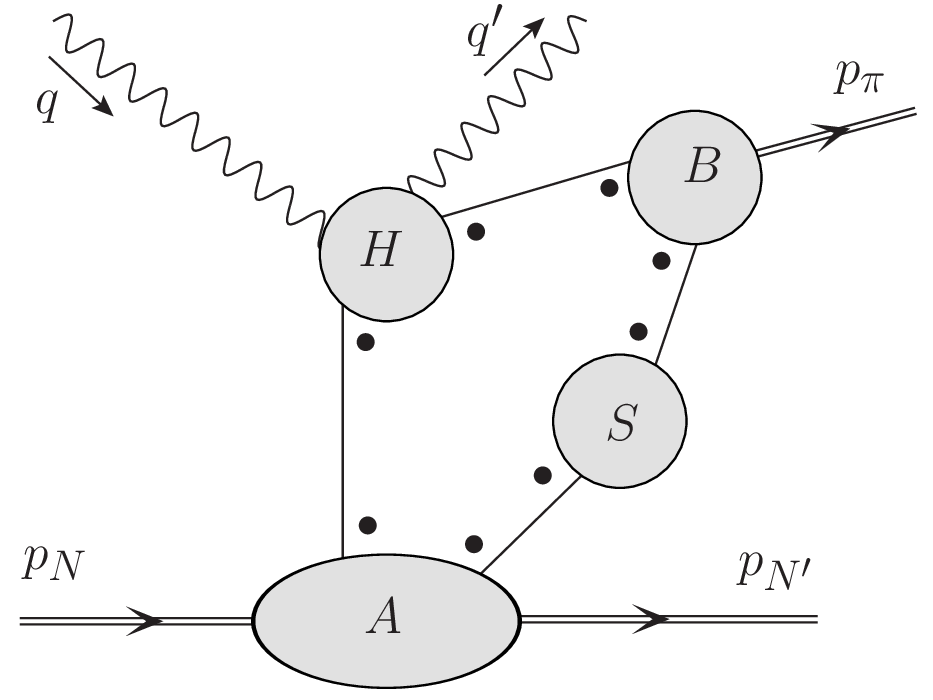}
(c)\hspace{4cm}(d)\hspace{0.4cm}
\vskip 5mm
\includegraphics[width=4.2cm]{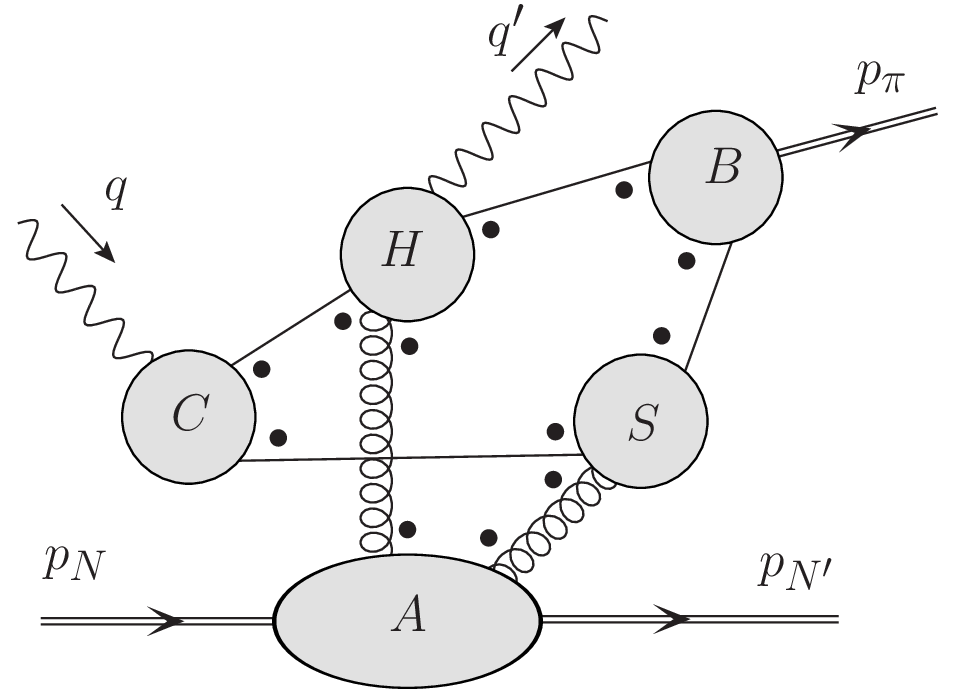}
\includegraphics[width=4.2cm]{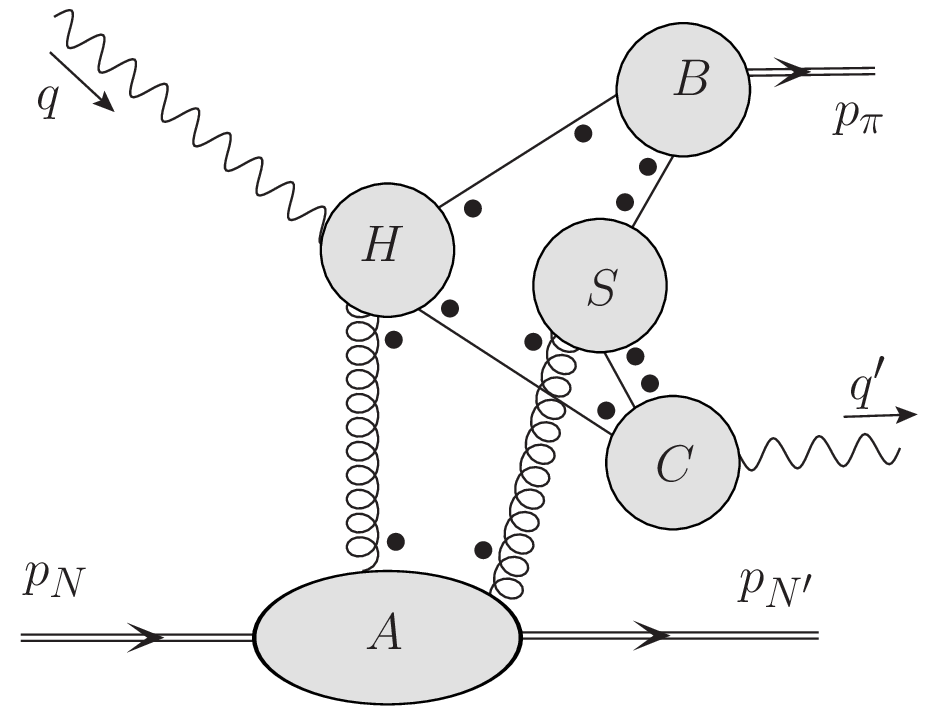}
(e)\hspace{4cm}(f)\hspace{0.4cm}
\caption{Superficially leading and super-leading reduced diagrams for the photoproduction of $  \pi ^{0}\gamma  $. The leading power is determined by graphs (a) and (b), which scale like $\lambda$. The $A/B$ subgraphs are collinear subgraphs in the nucleon/pion directions, $H$ is the hard subgraph and $S$ is the usoft subgraph. The $A$-to-$H$ gluons, which are shown in bold, are transversely polarised. The dots represent any number of $K$-gluons which accompany the line(s) in question.} 
\label{fig:reduced-diagrams}
	\end{figure}
In \FIG\ref{fig:reduced-diagrams}, we present a catalog of all relevant superficially leading and super-leading regions of the process in \EQ\eqref{eq:process-def}, in the spirit of the Coleman-Norton picture. ``Superficially'' means that the scaling obtained from the Libby-Sterman power counting in \EQ\eqref{eq:pc-formula} might still be corrected to be suppressed by powers of $\lambda$, due to cancellations when summing over all graphs. These possible cancellations, due to Ward identities (WIs), are a fundamental ingredient for factorization proofs in QCD. We discuss each of the reduced graphs in \FIG\ref{fig:reduced-diagrams}. 

We need to establish that there is no soft gluon which is trapped in the Glauber region. A necessary condition for the appearance of the Glauber pinch \footnote{See Chapter 5.6 of \cite{Collins:2011zzd}.} is the presence of at least two initial state external particles in different directions and two final external state particles in different directions, which are each connected to collinear subgraphs (given that there are no soft external particles). The external final state particles may or may not be in the same direction as the external initial state particles.

An example where a Glauber pinch exists, when the initial and final state collinear particles are in the same direction, is the double diffractive exclusive process $p_1 p_2 \to p_1' p_2' \gamma \gamma$, where $p_1/p_1'$ and $p_2/p_2'$ define two separate light-cone directions, and the two photons have large opposite transverse momenta, which forces them to be produced in the hard scattering. In this case, the trapped Glauber gluon is exchanged between the two collinear spectators in the light-like directions $p_1/p_1'$ and $p_2/p_2'$, respectively. 

On the other hand, if there is only one pair of external incoming and outgoing particles in the same direction, say $p_1/p_1'$ while $p_2$ and $p_2'$ are in different directions, then the Glauber gluon can be exchanged between a $p_1/p_1'$ collinear spectator and a soft spectator exchanged between $p_2$ and $p_2'$. This is the type of Glauber pinch that happens in the process we are considering here, and schematically corresponds to \FIG\ref{fig:reduced-diagrams}(e), with the $A$-to-$S$ gluon having Glauber scaling. The difference that a single photon is produced here, instead of two as in the previous example, is irrelevant for this discussion.

We remark that the above arguments imply that if incoming photon is virtual, then there cannot be a Glauber pinch. This is because, in that case, the incoming virtual photon has to be connected to the hard subgraph, and thus there cannot be more than one external incoming particle connected to collinear subgraphs.

\subsection{Graph (a)}

Consider first the reduced graphs in (a) of \FIG\ref{fig:reduced-diagrams}. In \cite{Qiu:2022pla}, it has been shown that these regions factorize in terms of a GPD and the pion DA. Standard arguments imply that the $K$-gluons from $A,B$ to $H$ subgraphs decouple after applying the region approximation when summing over $H$ subgraphs, forming the Wilson lines in the GPD and DA. Graph (a) scales as $\lambda^1$, which defines the \textit{leading} power for the process under consideration.

Furthermore, standard arguments show that the usoft subgraph decouples in (a),  and then gives unity. For this argument to work, it is necessary that an $A$-to-$B$ gluon\footnote{The subgraph $S$ can be `trivial', in the sense that it contains just a single line, such that an $A$-to-$S$-to-$B$ gluon is really just an $A$-to-$B$ gluon.} is not trapped in the Glauber region $l_g \sim (\lambda^2 ,\lambda^2 , \lambda)$. In \cite{Qiu:2022pla}, it was shown that for gluons connecting the $A$ and $B$ subgraphs, one can always deform the Glauber gluon to a collinear-to-$A$ gluon. We highlight that the necessary conditions for a Glauber pinch are not satisfied here, since the incoming photon is connected to the hard subgraph. Hence, there is only one external incoming particle connected to a collinear subgraph.

\subsection{Graph (b)}
For $t$-channel gluon exchanges, we explicitly consider separately the case where the $A$ and $H$ subgraphs are connected by two transversely polarised gluons, which is represented in graph (b). The situation when both or one of these two gluons are $K$-gluons corresponds to graph (c) instead, which is discussed in the next subsection.

The same arguments as in graph (a) apply here, which imply the standard factorisation in terms of a gluon GPD and DA for the pion, with the $K$-gluon factorised in terms of Wilson lines inside the GPD and DA, and the soft factor giving unity once again. The overall scaling of graph (b) is $\lambda^1$, like for graph (a).

\subsection{Graph (c)}
\label{sec:graph-c}
The exact power counting of graph (c) depends on whether the gluons connecting the subgraphs are $G$- or $K$-gluons. Na\"ively, the dominant power corresponds to $\lambda^{-1}$, which corresponds to the case where all the gluons are $K$-gluons.

First, we note that as for graphs (a) and (b), there is no Glauber pinch for (c), so the $A$-to-$S$ and $S$-to-$B$ gluons are strictly usoft.

We will now argue that, after the sum over all subgraphs, the contribution from graph (c) is actually power-suppressed, since there are at least three $G$-gluons (i.e. all the gluons that are explicitly drawn in graph (c) in \FIG\ref{fig:reduced-diagrams}).

Consider an $A$-to-$H$ gluon, which has a collinear momentum $l_c \sim (1, \lambda^2, \lambda)Q$. Then, the $K$ term is given by
\begin{align}
A_{\mu} \frac{n^{\mu} l_c^{\nu}}{l_c^+} H_{\nu}.
\label{eq:AH-K-term}
\end{align}
Next, consider an $S$-to-$B$ gluon, which has an ultrasoft momentum $l_s \sim (\lambda^2, \lambda^2, \lambda^2)$. The corresponding $K$  term reads
\begin{align}
S_{\mu} \frac{n^{\mu} l_s^{\nu}}{l_s^+} B_{\nu}.
\label{eq:SB-K-term}
\end{align}
For what follows, it will be useful to group certain subgraphs together - Thus, we denote the grouping of the subgraphs $X_1$ and $X_2$ by $X_1 \times X_2$. It is apparent that the $K$ terms of the gluons connecting $A \times S$ to $H \times B$ are proportional to $l \cdot  (H \times B)$. In an Abelian gauge theory, $l \cdot \sum (H \times B) = 0$ where the sum is over all $H \times B$ subgraphs, so that the $K$ components vanish.\footnote{The sum over graphs also includes $S$-to-$H$ attachments, but these are power-suppressed by \EQ\eqref{eq:pc-formula-2}, so they do not influence the argument.}

In QCD, this is more involved due to its non-Abelian structure. However, it is well-established that gauge invariance dictates that, for a group of one or more subgraphs that have no quark/antiquark attachments, at least two gluon attachments must be $G$-gluons in order not to get zero.\footnote{This is discussed in Chapter 11.9 of \cite{Collins:2011zzd}.} The remaining gluons can be either $G$- or $K$-gluons - However, having more than two $G$-gluons leads to even further power suppression. Thus, we need to have at least two $G$-gluons between $\sum H \times B$ and $\sum A \times S$.

Moreover, we can apply the same argument for the gluons between $\sum H \times B \times S$ and $\sum A$. The corresponding $\bar{K}$ term of an $S$-to-$A$ gluon reads
\begin{align}
S_{\mu} \frac{\bar n^{\mu} l_s^{\nu}}{l_s^-} A_{\nu}.
\label{eq:SA-K-term}
\end{align}
Using the same line of thought as before, this implies that we need to have at least two $G$-gluons between $\sum H \times B \times S$ and $\sum A$. Since we have already established that there needs to be at least two $G$-gluons between $\sum H \times B$ and $\sum A \times S$, the consequence is that there needs to be at least three $G$-gluons in total.\footnote{The same argument can be used to argue that there exists a contribution with two $A$-to-$H$ $G$-gluons. This corresponds exactly to graph (b).} Hence, the power counting of graph (c) is actually $\lambda^2$.

A situation with exactly three $G$-gluons corresponds to graph (c) where all the explicitly drawn gluons are $G$-gluons. We note that other situations may occur, for instance, having two $A$-to-$S$ $G$-gluons and two $S$-to-$B$ $G$-gluons. However, the latter situation suffers from even further suppression, by one extra power of $\lambda$.

\subsection{Graph (d)}
\label{sec:graph-d}

This region has already been mentioned in \cite{Qiu:2022pla}. Superficially, this region is of the leading power (i.e. it scales as $\lambda^1$). The additional assumption that is made relies on the so-called ``soft-end suppression'', which states that the leading twist pion DA $\phi_{\pi}(z)$ vanishes linearly at the endpoints $z = 0,1$. This fact is well-established \cite{Ball:1998je}. This implies that the $B$ subgraph, when integrated over $k^+$ and $k_{\perp}$, behaves like $k^-$ as $k^- \rightarrow 0$. Therefore, we get a further suppression compared to \EQs\eqref{eq:pc-formula} and \eqref{eq:pc-formula-2}, making graph (d) power-suppressed. In \SEC\ref{sec:soft-end-suppression}, we provide some more details on this issue.

Note that in the related process of DVMP, it is not necessary to invoke soft-end suppression. This is because when the incoming virtual photon is longitudinally polarised, one automatically obtains additional suppression from the hard subgraph \cite{Collins:1996fb}. However, this argument does not work here, so one needs to rely on soft-end suppression for exclusive $\pi^0\gamma$ photoproduction.

\subsection{Graph (e)}
\label{sec:graph-e}

Superficially, it scales as $\lambda^0$. However, by the same argument as in \SEC\ref{sec:graph-c}, using the WI, at least two of the gluons connecting the collinear subgraph $A$ with $\sum H \times B \times S \times C$ must be $G$-gluons. This introduces a suppression of $\lambda^2$, bringing the effective scaling of graph (e) to be $\lambda^2$. On top of this, graph (e) experiences soft-end suppression.

Still, it should be noted that this argument applies only if all lines in the $S$ subgraph are strictly usoft. As mentioned earlier, graph (e) corresponds to a situation where one can have a Glauber gluon exchange between a collinear spectator inside subgraph $A$ and the soft quark spectator connecting collinear subgraphs $B$ and $C$ (which passes through subgraph $S$). We will later show through an explicit example in \SEC\ref{sec:example} that we have a pinched configuration when an $S$-to-$A$ gluon has the Glauber scaling $l \sim (\lambda , \lambda^2 ,\lambda)$,  while the $S$-to-$B$ and $S$-to-$C$ quark lines have the soft scaling $k \sim (\lambda, \lambda, \lambda)$. It is worth highlighting that the Libby-Sterman power counting rules in \EQs\eqref{eq:pc-formula} and \eqref{eq:pc-formula-2} do not apply for these scalings. Consequently, we perform an explicit power counting of this pinched configuration in \SEC\ref{sec:pc-glauber}, and we find that it scales as $\lambda$, which is the leading power defined by graphs (a) and (b). Furthermore, we stress that soft-end suppression is already taken into account by our power counting, such that it is not enough to make this region power-suppressed.

For the Glauber gluon $l\sim (\lambda , \lambda^2 ,\lambda)$, the key observation is that 
\begin{align}
l \cdot A &\sim l^- A^+ + l_{\perp} \cdot A_{\perp} ,
\\
l \cdot S &\sim l^+ S^- + l_{\perp} \cdot S_{\perp}.
\end{align}
In particular, $l^- A^+ \sim l_{\perp} \cdot A_{\perp}$ and $l^+ S^- \sim l_{\perp} \cdot S_{\perp}$. The contributions $l_{\perp} \cdot A_{\perp}$ and $l_{\perp} \cdot S_{\perp}$ correspond to the $G$ term, which is not suppressed compared to the $K$ term. Thus, the $G$ term ``survives'' after using the WI, and the overall power for this region remains $\lambda$.
This is discussed in detail in an explicit example in \SEC\ref{sec:WIs}. 

Therefore, collinear factorization is broken for the exclusive photoproduction of $\pi^0 \gamma$ pair, since there is a Glauber pinch of the $A$-to-$S$ gluon in graph (e), which contributes to the leading power as defined by graphs (a) and (b).

\subsection{Graph (f)}
\label{sec:graph-f}
For completeness, we also consider the graph (f), which at first sight is similar to (e), in that it has a subgraph $C$ collinear to an external photon. As before, when the $S$ subgraph only involves usoft momenta, the same argument as for graph (e) applies, such that we find that the overall scaling is $\lambda^2$, which is a suppressed contribution.

We highlight that there is no Glauber pinch for this configuration, since there is only one initial state particle connected to a collinear subgraph, here $A$ (the incoming photon is connected to the hard subgraph $H$). Thus, the necessary condition for a Glauber pinch to exist, discussed earlier, is not satisfied here.

\section{Explicit example with Glauber pinch}
\label{sec:example}

As explained in \SEC\ref{sec:graph-e}, collinear factorisation for the exclusive photoproduction of a $\pi^0 \gamma$ pair is broken by the presence of a Glauber pinch, which we can identify with \FIG\ref{fig:reduced-diagrams}(e) when the $A$-to-$S$ gluon has Glauber scaling. To be clear, the region that actually ``breaks'' factorization is not the case where the lines of the $S$ subgraph are usoft, but are instead pinched in a more complicated configuration that prevents the relative suppression of the $G$-gluons. Therefore, throughout this section, whenever we refer to graph (e) in \FIG\ref{fig:reduced-diagrams}, we specifically mean the Glauber pinch configuration.

\begin{figure}
		\centering
		\includegraphics[width=9cm]{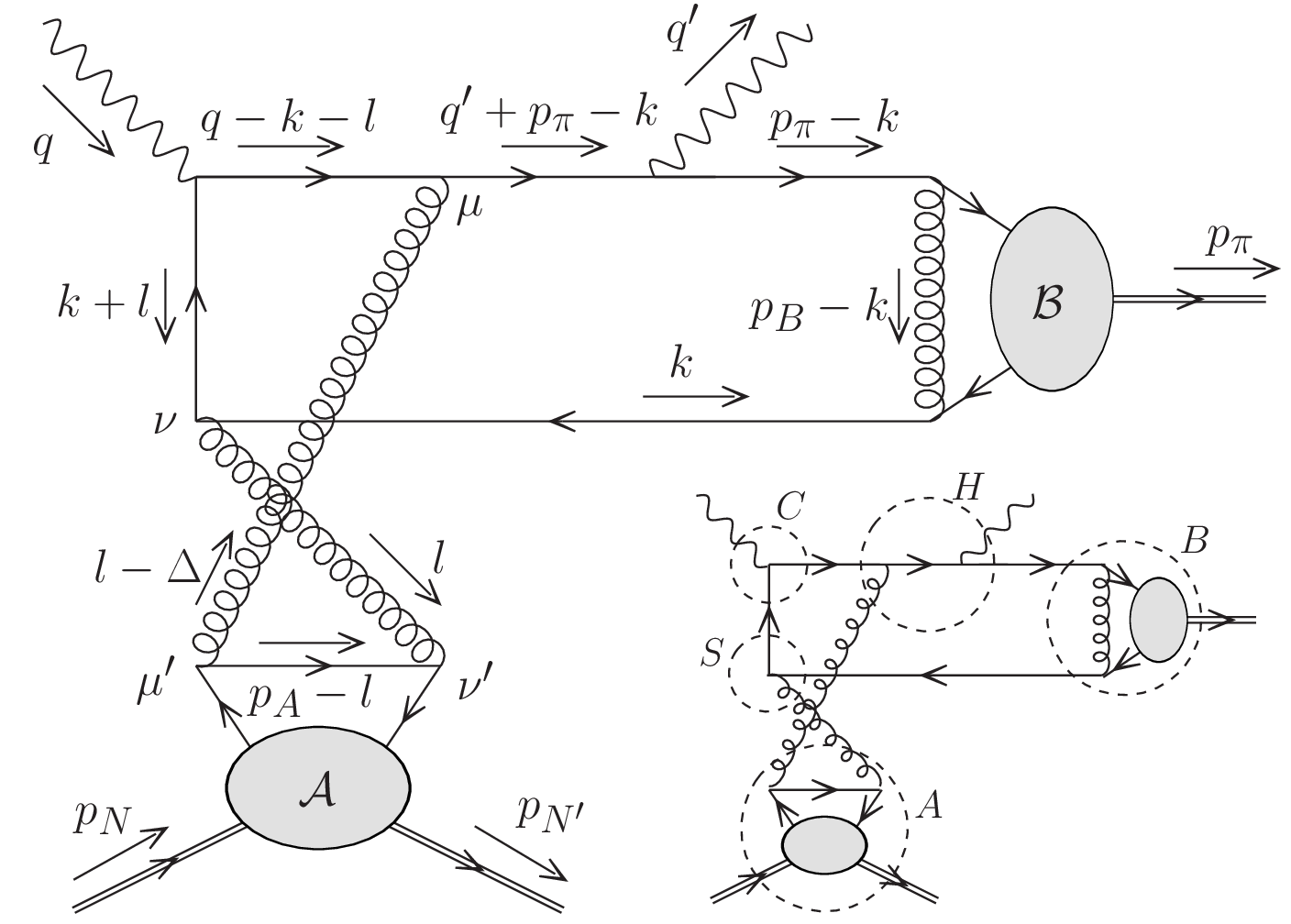}
		\caption{A diagram that has a Glauber pinch. Bottom right: For the Glauber scaling in \EQ\eqref{eq:Glauber-scaling}, it corresponds to \FIG\ref{fig:reduced-diagrams}(e).}
		\label{fig:explicit-2-loop}

	\end{figure}

In this section, we consider explicitly the two-loop diagram in \FIG\ref{fig:explicit-2-loop}, which has a region of loop momentum space that corresponds to graph (e). This is similar to the type of analysis that was done in the factorization proofs for DVCS \cite{Collins:1998be} and DVMP \cite{Collins:1996fb}.

	We recall that the scalings for the external particles are given in \EQs\eqref{eq:external-scalings-nucleon} and \eqref{eq:external-scalings-pion}. The blobs $ {\cal A} $ and ${\cal B}$ contain only lines that are collinear to the nucleon and pion systems respectively. We will use the fact that they are proportional to the unit matrix in color space, which is true because the external particles are color singlet states. In the blob ${\cal A} $, the momenta are dominated by the collinear direction along $ p_{N} $ and $ p_{N'} $, while in blob ${\cal B} $, the momenta are proportional to the pion momentum $ p_{\pi} $. The momenta $p_A \sim (1,\lambda^2,\lambda)$ and $p_B \sim (\lambda^2, 1, \lambda)$ are loop momenta that circulate only through the subgraphs $A$ and $B$ respectively.
 
 For our purposes, it is sufficient to describe the sub-amplitudes ${\cal A}$ and ${\cal B}$, which we define to include the external quark propagators as well as $d^4p_A$ and $d^4p_B$ respectively, by their vector component in Dirac space.\footnote{Strictly speaking, one should project onto the axial-vector contribution of ${\cal B}$ for the leading twist pion DA, but we consider the vector case here for simplicity.} That is, they can be taken to be ${\cal A}^{\mu} \gamma_{\mu}$ and ${\cal B}^{\mu} \gamma_{\mu}$ respectively, where, by dimensional analysis and Lorentz covariance, we have ${\cal A} \sim (1, \lambda^2, \lambda)$ and ${\cal B} \sim (\lambda^3, \lambda, \lambda^2)$. To see this, we can use the argument from Chapter 5.5 of \cite{Collins:2011zzd}: Consider the sub-amplitude ${\cal A}^{\mu}$ in the rest frame of the nucleon. By dimensional analysis (since $Q=1$), ${\cal A}^\mu_{\mathrm{RF}} \sim (\lambda,\lambda,\lambda)$. Now, we boost to the frame where $p_N,\, p_{N'}$ have large momentum in the $\bar{n}$-direction. By Lorentz covariance, after the boost, we can write ${\cal A}^{\mu} \sim \lambda^{a} p^{\mu}$, where $a$ is some number. To fix it, we note that the $\perp$ component of ${\cal A}^\mu$ remains unchanged after the boost. Therefore, we conclude that $a = 0$, which means ${\cal A}^{\mu} \sim  p^{\mu} \sim (1, \lambda^2, \lambda)$, as was claimed. We note that this analysis works since the sub-amplitude ${\cal A}^\mu$ contains only lines with momenta that have $\bar{n}$-collinear scaling.  A similar analysis is also performed for $\cal{B}$. The additional power of $\lambda$ in ${\cal B}_{\perp}$ compared to ${\cal A}_{\perp}$ is due to the fact that the subgraph ${\cal B}$ has one external line less.

	To derive the scaling, we write the diagram in \FIG\ref{fig:explicit-2-loop}, using the Feynman gauge, as
	\begin{align}
		\label{eq:2-loop-amplitude}
	\int F_{A}^{ \mu ' \nu '} g_{ \mu  \mu '}g_{ \nu  \nu '} \tr  \left[   {F}_{H}^{ \mu  \nu } {F} _{B}  \right] \;,
	\end{align}
	where, omitting the $i\epsilon$ prescriptions,
	\begin{align}
 \label{eq:FA}
		F_{A}^{ \mu'  \nu '} & =  dl^{-}d^{2}l_{\perp}  \left( \frac{\tr \left[  {\slashed{\cal A}}\gamma ^{ \nu '}  \left( \slashed{p}_A- \slashed{l} \right)  \gamma ^{ \mu '}     \right] }{l^2  \left( l- \Delta  \right)^2  \left( p_A-l \right)^2  } \right)\;, \\[5pt]
		F_{B} & = dk^{+}d^{2}k_{\perp} \left( \frac{ \slashed{k} \slashed{\cal B} \left(  \slashed{p}_{ \pi }- \slashed{k}   \right)   }{k^2  \left( p_{ \pi }-k \right)^2  \left( p_B-k \right)^2  } \right) \;,   \label{eq:FB} \\[5pt]
		F_{H}^{ \mu  \nu }& = dk^{-}dl^{+}  \label{eq:FH} \\ & \times \frac{ \slashed{\epsilon }_{q'}^{*} \left(  \slashed{q}'+ \slashed{p}_{ \pi }- \slashed{k}    \right)\gamma ^{ \mu } \left(  \slashed{q}- \slashed{k}- \slashed{l}    \right) \slashed{\epsilon }_{q} \left(  \slashed{k}+ \slashed{l}   \right) \gamma ^{ \nu }   }{ \left( q'+p_{ \pi }-k \right)^2  \left( q-k-l \right)^2  \left( k+l \right)^2   } \;. \nonumber
	\end{align}
 The grouping of the measures of the loop momentum components in the above expressions (relevant for the scaling) has been done in the spirit of collinear factorization. That is, with the collinear scaling, $F_A$ can be ``identified'' with the GPD, $F_B$ with the pion DA and $F_H$ with the coefficient function. To consider the most general case, however, we keep \textit{all} the integral signs outside the different parts on the RHS of \EQ\eqref{eq:2-loop-amplitude}.

 \subsection{Collinear pinch}
 \label{sec:coll pinch}

We start by reviewing the familiar case of the collinear pinch, which corresponds to the region in \FIG\ref{fig:reduced-diagrams}(b) of \FIG\ref{fig:explicit-2-loop}. The pinching of the quark of momentum $k$ in the region $k \sim (\lambda^2 , 1 , \lambda)$ follows immediately from the analysis in \SEC\ref{sec:bubble-diag-pinch}.
For the $l$ momentum, the relevant propagators are, assuming that $l_{\perp} \sim \lambda$,
\begin{align} \notag
l^2 + i\epsilon &= 2l^+ l^- + O(\lambda^2) + i\epsilon,
\\
(l- \Delta)^2 +i \epsilon &= 2 (l^+ - \Delta^+) l^- + O(\lambda^2) + i\epsilon,
\label{eq:l denoms}
\\ \notag
(l - p_A)^2 + i\epsilon &= 2 (l^+ - p_A^+) l^- + O(\lambda^2) + i\epsilon.
\end{align}
It is clear that $l^-$ is pinched to be of  $O(\lambda^2)$ in various regions of $l^+ \sim \lambda^0$
depending on the sign of $p_A^+ \sim \lambda^0$ and $\Delta^+ \sim \lambda^0$.
Indeed, one can readily show that the conditions for the pinch are
\begin{align}
\max(\Delta^+, p_A^+, 0) > l^+ > \min(\Delta^+, p_A^+, 0).
\label{eq:cond1}
\end{align}
This can be seen either directly, by observing whether there are poles on opposite sides of the $l^-$ contour, or by using the Landau condition in the limit $\lambda \rightarrow 0$, see \APP\ref{app:collinear-pinch-Landau}. One must be careful, however, whenever the plus momenta of the lines involved become much smaller than unity. In that case, the collinear scaling $(1, \lambda^2, \lambda)$ is not correct and the corresponding line should rather be described by a (ultra)soft scaling $(\lambda_s, \lambda_s, \lambda_s)$, which we treat separately. 

\begin{figure}
    \centering
    \includegraphics[width=8.5cm]{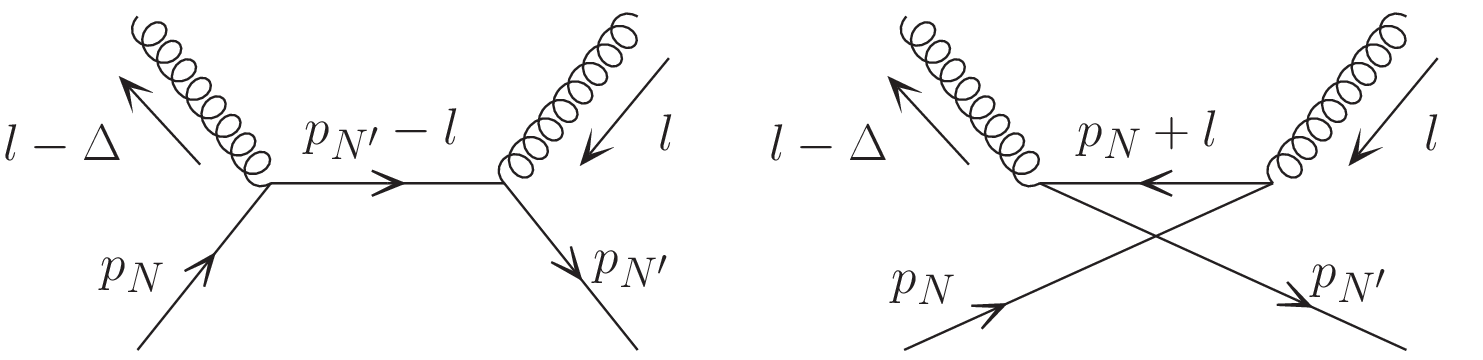}
    \caption{Diagrams which illustrate the collinear pinch in the trivial case, where the target is just a quark.}
    \label{fig:collinear-pinch-propagators}
\end{figure}

Since kinematically $\Delta^+ < 0$, \EQ\eqref{eq:cond1} becomes
\begin{align} \notag
p_A^+ > l^+ > \Delta^+, \qquad & p_A^+ > 0,
\\
0 > l^+ > \min(\Delta^+, p_A^+),  \qquad &p_A^+ < 0. \label{eq:ranges}
\end{align}
Thus, we conclude that $l$ is trapped in the collinear region $l \sim (1, \lambda^2, \lambda)$. 
We use that to justify inductively that $p_{N'}^+ \geq p_A^+ \geq - p_N^+$. For example, if $\mathcal A$ is trivial in the sense that we consider the target to be just a quark, so that it consists of the sum of the two diagrams in \FIG\ref{fig:collinear-pinch-propagators}, we can have either $p_A^+ = p_{N'}^+ = (1-\xi) P^+ > 0$ for the uncrossed and $p_A^+ = - p_N^+ = - (1+\xi)P^+ < 0$ for the crossed graph. Identifying $l^+ = (x-\xi)P^+$, \EQ\eqref{eq:ranges} results in the well-known classifications of the DGLAP and ERBL region:
\begin{align} \notag
1 > x > +\xi, \qquad &\text{(DGLAP I)}
\\
+\xi > x > -\xi, \qquad &\text{(ERBL)}
\\ \notag
-\xi > x > -1. \qquad &\text{(DGLAP II)}
\end{align}

It is interesting to remark that the $l^2 + i\epsilon$ is not needed to pinch $l^- = O(\lambda^2)$ in the region $p_{N'}^+ > l^+ > \Delta^+$. Hence, $l^+$ can actually be much smaller than $\lambda^0$, i.e. close to the boundary between ERBL and DGLAP regions, without ``losing'' the pinch in $l^-$. Indeed, the corresponding pole for $l^-$ from the $l^2 + i\epsilon$ propagator is $l^- = - \frac{l_{\perp}^2}{2l^+}$, which is separated from the origin by a distance much larger than $\lambda^2$. However, the pinch that forces $l^- = O(\lambda^2)$ is retained by virtue of the other two propagators in \EQ\eqref{eq:l denoms}. Note that in this case, the $l^2 + i\epsilon$ denominator becomes ``off-shell'' in the sense that $l^- \sim \lambda^2 \ll | \frac{l_{\perp}^2}{2l^+}|$, which implies that $l$ has a Glauber scaling, see \EQ\eqref{eq:glauber-defintion}. 
This is an important feature of the off-forward kinematics (since $\Delta^+ < 0$), which will be relevant in the next subsection.

	\subsection{Pinch Analysis for the Glauber region}

	\label{sec:pinch-analysis}
\begin{figure}
    \centering
    \includegraphics[scale=0.5]{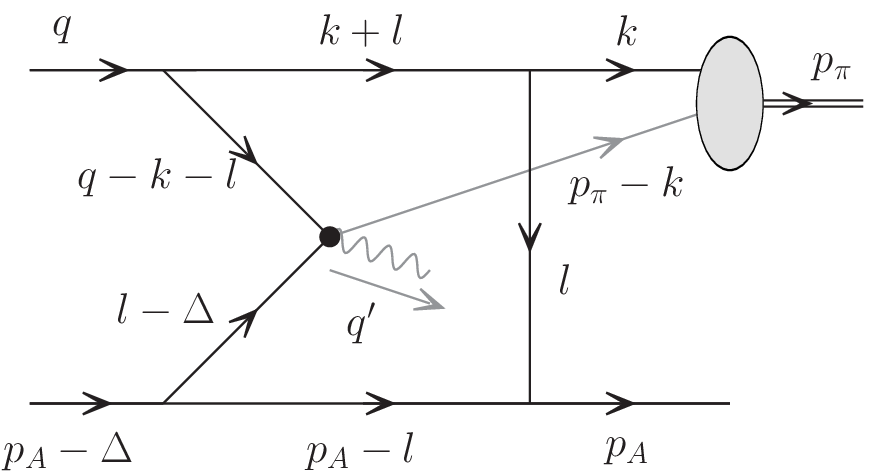}
    \caption{The simplified version of \FIG\ref{fig:explicit-2-loop}, where the arrows indicate the flow of internal loop momenta. For the Glauber pinch analysis, it is sufficient to focus on the propagators that depend on $l$, and not $k$. The lines of momentum $p_{\pi} - k$ and $q'$, drawn in grey, are not relevant for the Glauber pinch analysis. The black dot denotes the hard reduced vertex.}
    \label{fig:glauberoneloop}
\end{figure}

From the discussion at the beginning of \SEC\ref{sec:reduceddias}, we have established that the soft pinch is always present, and should be considered. Therefore, we start our analysis in this section by considering the scalings $k \sim (\lambda, \lambda, \lambda)$ and $l \sim (\lambda, \lambda, \lambda)$. However, due to the poles of the other propagators in the amplitude, it might be that the $+$ and/or $-$ components are pinched to be much smaller, e.g. $\lambda^2$, while the transverse component still scales as $ \lambda$. This is precisely the Glauber pinch situation. The crucial question then is whether in the neighborhood of the soft pinch, there are poles which force $l$ to be in a Glauber-like region in the diagram in \FIG\ref{fig:explicit-2-loop}.

As argued above, we fix $k \sim (\lambda,\lambda,\lambda)$, and $l_{\perp}\sim \lambda$, and focus on the pole structure of $l^{\pm}$. The pinch analysis then reduces to the one-loop graph in \FIG\ref{fig:glauberoneloop}. Recall that $\Delta^+ < 0$ and $q^- > 0$ kinematically. 
Then, the four relevant propagators can be written as
\begin{align}
\label{eq:pA-l}
(p_A - l)^2 + i \epsilon &= 2p_A^+ (-l^- + O(\lambda^2) + \text{sgn}(p_A^+) i\epsilon), 
\\
\label{eq:Delta-l}
(\Delta - l)^2 + i \epsilon &= - 2 \Delta^+ (l^- + O(\lambda^2) + i \epsilon),
\\
\label{eq:q-k-l}
(q-k-l)^2 + i \epsilon &= 2q^- (- l^+ + O(\lambda) +  i\epsilon),
\\
\label{eq:k-plus-l}
(k+l)^2 + i \epsilon &= 2k^- (l^+ + O(\lambda) + \text{sgn}(k^-) i\epsilon).
\end{align}
It can be easily checked that these estimates remain true also in the center-of-mass frame wrt to $q$ and $\Delta$, where $q$ has its dominant component along $n$, and $p_A$ and $\Delta$ have their dominant components along $\bar n$.

From \EQs\eqref{eq:pA-l} and \eqref{eq:Delta-l}, we find that $l^{-}$ is pinched to be of $O(\lambda^2)$ for $p_A^+>0$. This was also identified in our analysis in \SEC\ref{sec:coll pinch}, and this fact alone does not imply that $l$ in pinched in the Glauber region. 
We now need to also check whether $l^+$ can be deformed to be much larger than $\lambda$. From \EQs\eqref{eq:q-k-l} and \eqref{eq:k-plus-l}, we find that $l^{+}$ is pinched\footnote{In the case where the incoming photon is virtual, the $(q-k-l)^2$ propagator is hard by definition, so $l^+$ is not pinched anymore. Hence, the corresponding Glauber pinch, identified here in photoproduction, does not occur.} to be of $O(\lambda)$ for $k^{-}>0$.

It is important to highlight that the pinches in $l^+$ and $l^-$ here do not depend on the $l^2 + i \epsilon$ propagator, which is actually off-shell here. Instead, they depend only on the four equations in \eqref{eq:pA-l} to \eqref{eq:k-plus-l}. This is typical of Glauber scalings. Moreover, we note that it is precisely the fact that we are in off-forward kinematics that allows the Glauber pinch here to occur \textit{only} from the two gluons probed from the nucleon, which are usually the ``active'' partons that participate in the scattering. This is in stark contrast with Glauber pinches that occur in cases with forward kinematics, for example in Drell-Yan, where the Glauber gluon is never one of the ``active'' partons.

We thus observe that $l^+$ is pinched at the incoming photon vertex with the poles being separated by distance $\sim \lambda$, implying that $l^+ \sim \lambda$ has the size a soft scale. On the other hand, due to the different signs of $\Delta^+$ and $p_A^+$, $l^- \sim \lambda^2$ is still pinched to be the size of the minus component of a collinear momentum, i.e. the usoft scale. 

An important point is that the Glauber pinch might be ``spurious" in the sense that it disappears after translating the loop momenta in a certain way. This argument can be crucial in showing that the Glauber pinches are absent, for example in diffractive DIS (see \SEC III E of \cite{Collins:1997sr}). Thus, it is necessary to show that the Glauber pinch is there for all possible routings of loop momenta.
In our case, we could think of translating $k \rightarrow k - l$, which reroutes the $l$ loop momentum through the meson lines. However, it is easy to see in that case that the lines carrying momenta $p_{\pi} - k + l$ and $k - l$ still pinch $l^+$ to be $O(\lambda)$. For the case where we swap $q \leftrightarrow q'$ in \FIG\ref{fig:explicit-2-loop}, it can be shown that there always exists a rerouting of the loop momentum $l$ such that there is no Glauber pinch. This confirms our initial statement regarding the necessary conditions to have a Glauber pinch in \SEC\ref{sec:reduceddias}. This important point is discussed in detail in \SEC\ref{sec:rerouting-momenta}.

\begin{figure}
    \centering
    \raisebox{0.5cm}{\includegraphics[scale=0.5]{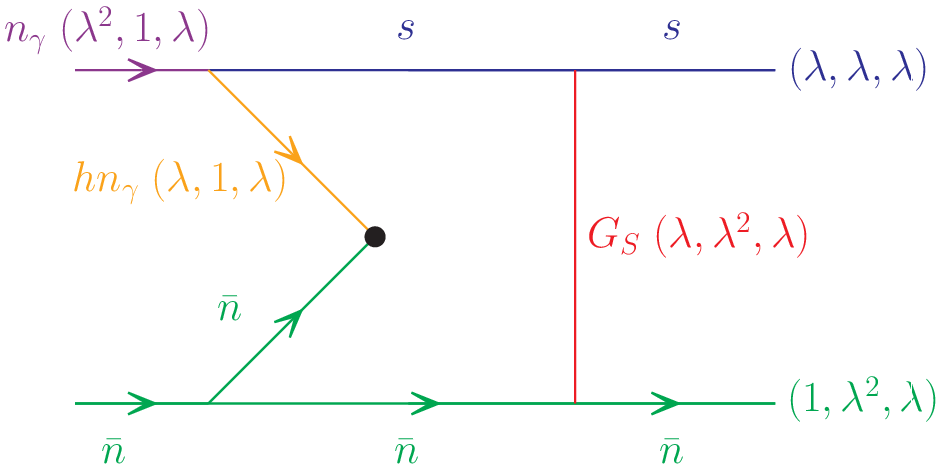}}
    \includegraphics[scale=0.5]{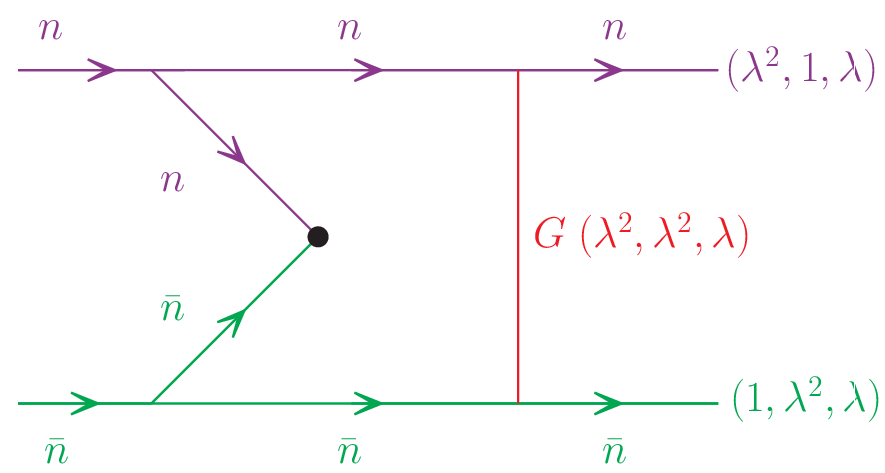}
    \caption{Comparison between the Glauber pinch in our case and the Drell-Yan process. Top: Collinear-to-soft Glauber exchange in the CM frame wrt $q$ and $\Delta$. This corresponds to our case, c.f.~\FIG\ref{fig:glauberoneloop} with the grey lines removed. Bottom: Collinear-to-collinear Glauber exchange. This corresponds to the Drell-Yan case. In both cases, the symbols next to the lines denote the scaling of the momentum of that line, which is $\bar n$($n$) for $\bar n$($n$)-collinear shown in green (magenta), $s$ for soft (blue), $n_{\gamma}$ for collinear to the incoming photon (magenta) and $hn_{\gamma}$ for hard-collinear to the incoming photon (orange). Here, ``hard-collinear'' means that the component along the direction of the incoming photon is of order $1$ while all other components are of order $\lambda$. The black dot represents the hard vertex, where all the hard propagators have been shrunk. The symbol $G$ ($G_S$) corresponds to a collinear-to-collinear (collinear-to-soft) Glauber gluon. Note  that we take the initial photon to be quasi-real for the sake of generality.}
    \label{fig:comparison-with-DY}
\end{figure}

Therefore, we conclude that $l$ is trapped in the $\bar{n}$-coll.~to soft Glauber region $l \sim (\lambda, \lambda^2, \lambda)$.
Note that the topology of the graph in \FIG\ref{fig:glauberoneloop} is actually very similar to the classic Glauber pinch that occurs in the Drell-Yan process.\footnote{We remark the well-known fact that in the inclusive Drell-Yan process, the Glauber contribution, while pinched, is canceled at the cross-section level due to unitarity.} In upper diagram of \FIG\ref{fig:comparison-with-DY}, the Glauber exchange occurs between the spectator collinear quark of the nucleon, and the \textit{soft} spectator quark joining the incoming photon with the  outgoing meson. This corresponds to a $\bar{n}$-coll.~to soft Glauber, \EQ\eqref{eq:nbar-coll-to-soft-Glauber}, which is exactly the pinch that we demonstrate in this subsection. In the Drell-Yan case, on the other hand, the Glauber exchange occurs between two collinear spectator partons of the two incoming nucleons, as shown in the lower diagram in \FIG\ref{fig:comparison-with-DY}. This leads to the `standard' collinear-to-collinear Glauber scaling in \EQ\eqref{eq:coll-to-coll-Glauber}.

\subsection{Rerouting of momenta in pinch analysis}

\label{sec:rerouting-momenta}

\begin{figure}
    \centering
    \includegraphics[width=7cm]{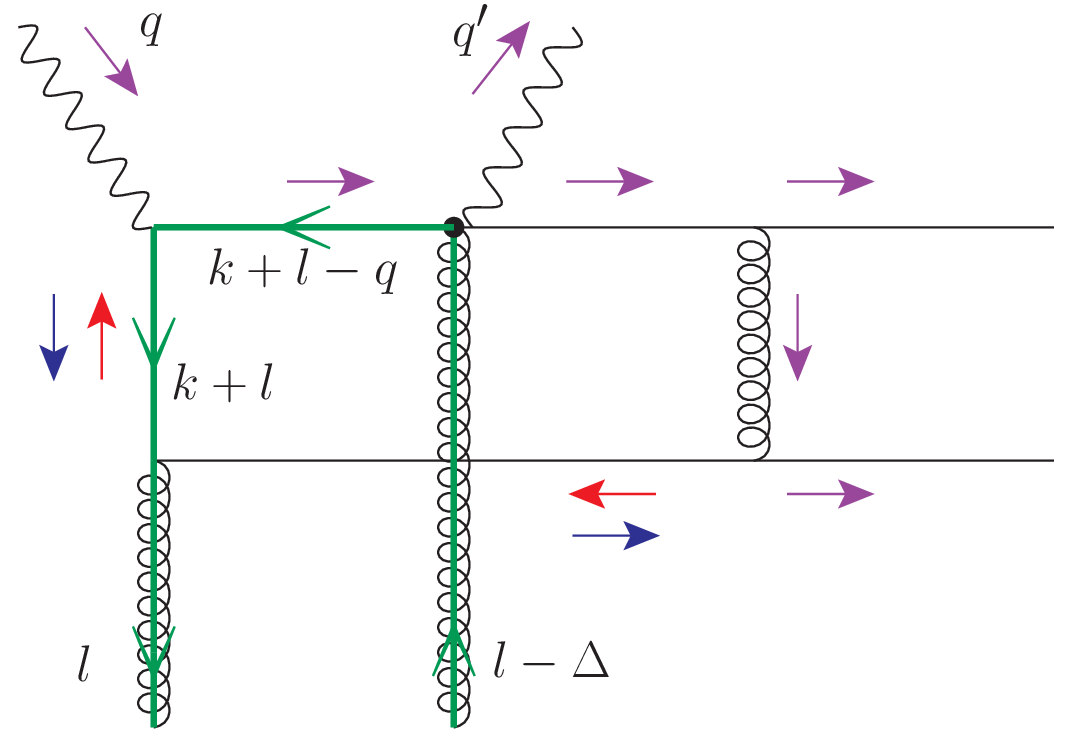}
        \\[15pt]
    \includegraphics[width=7cm]{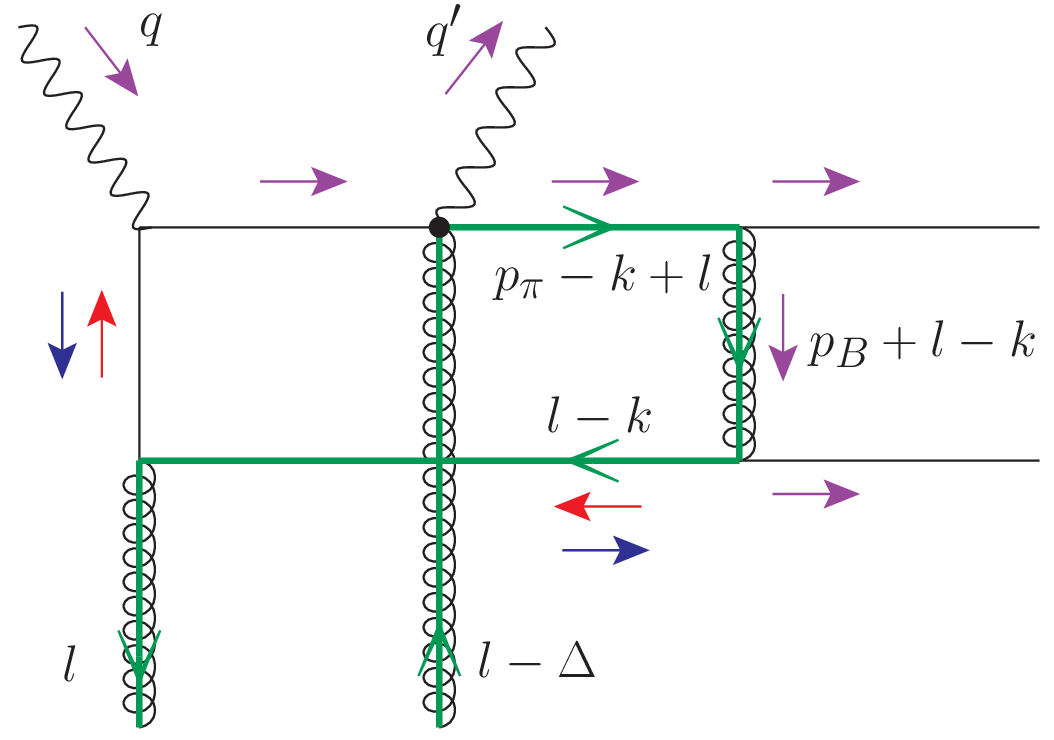}
    \caption{Diagram (e) of \FIG\ref{fig:reduced-diagrams}, with two possible routings of the $l$ momentum, shown by the green arrows on the lines. The labels next to the lines correspond to the momenta in the direction of the green arrows. The purple arrows show the flow of large $\sim \lambda^0$ $n$-momentum component. The blue (red) arrows indicate the flow of small $\sim \lambda^1$ $n$-momentum component for $k^->0$ ($k^-<0$).  $l^+$ is pinched in the case of $k^->0$.}
    \label{fig:momentum-rerouting-diag-e}
\end{figure}

\begin{figure}
    \centering
    \includegraphics[width=7cm]{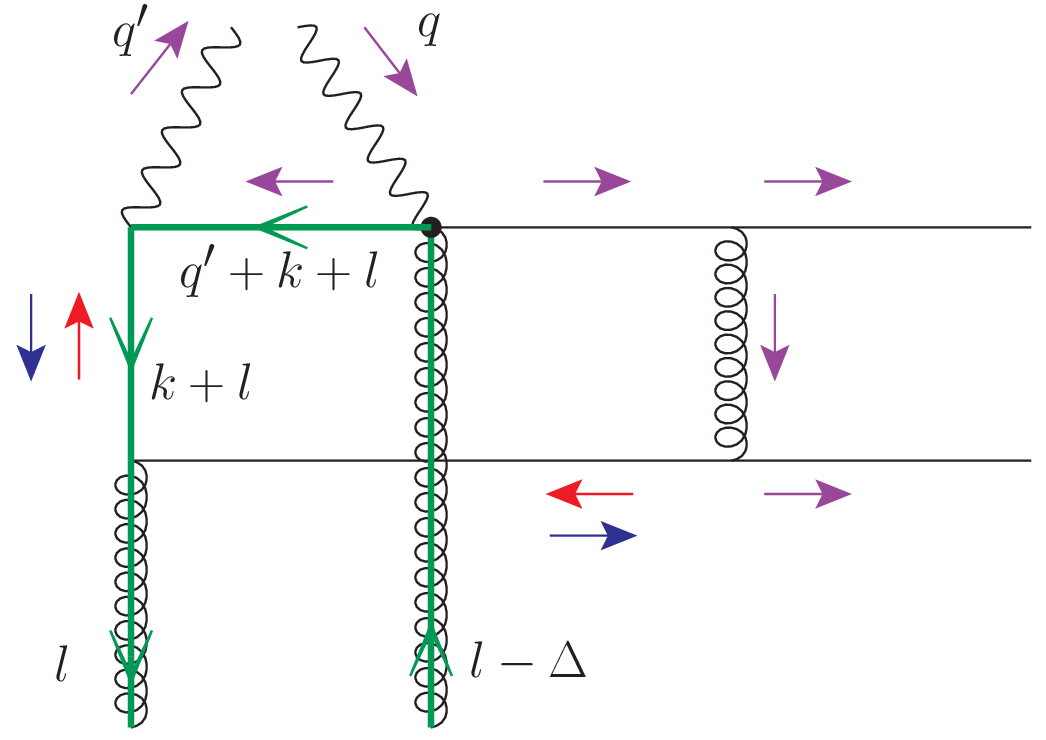}
    \\[15pt]
    \includegraphics[width=7cm]{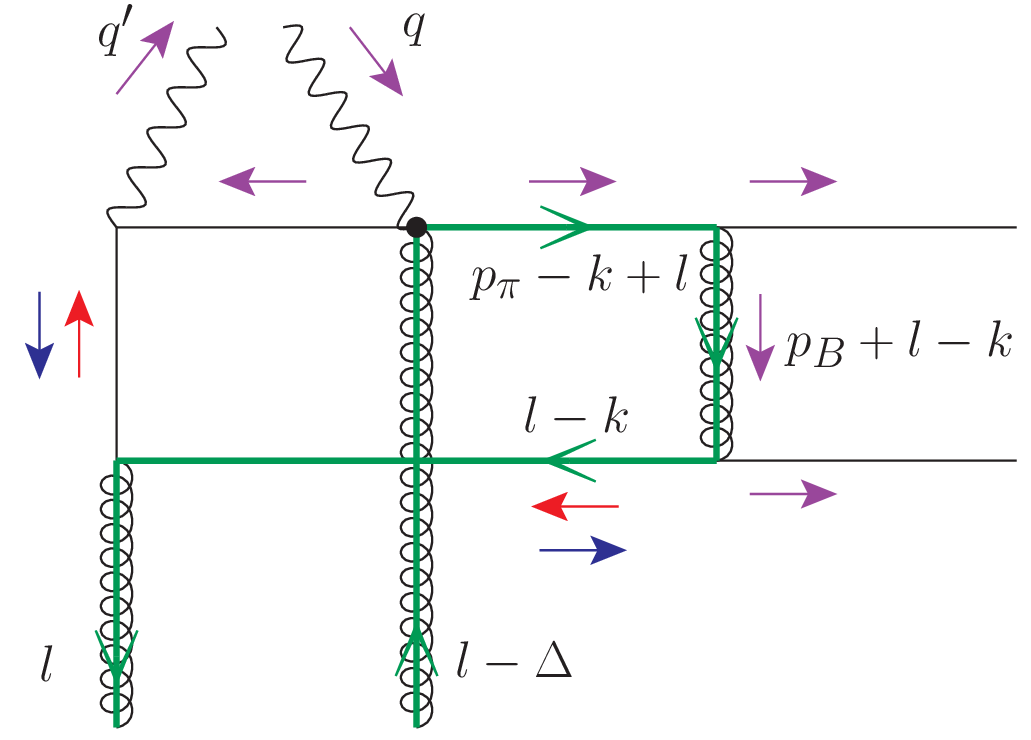}
    \caption{Diagram (f) of \FIG\ref{fig:reduced-diagrams}, with two possible routings of the $l$ momentum. The same notation as in \FIG\ref{fig:momentum-rerouting-diag-e} is used here. In this case, there is no ``true'' pinch in $l^+$, no matter what the sign of $k^-$ is.}
    \label{fig:momentum-rerouting-diag-f}
\end{figure}

To show that a given diagram suffers from a Glauber pinch, it is not sufficient to show that such a pinch takes place for a particular routing of the loop momentum. Indeed, there generically exist such ``spurious'' Glauber pinches which happen if the loop momentum that is supposedly pinched in the Glauber region is routed in a ``bad'' way. In other words, while a Glauber pinch might be there for some particular routing, it might disappear if one translates some other loop momentum in a particular way. This argument is crucial in showing the absence of Glauber pinches in diffractive hard scattering \cite{Collins:1997sr}. 

Thus, in order to prove the claim that the process considered here suffers from a Glauber pinch, it remains to show that the Glauber pinch persists if one routes the loop momentum $l$ in \FIG\ref{fig:explicit-2-loop} the other way through the meson system,\footnote{Note that there are only two possible routings for the momentum $l$, since our explicit example in \FIG\ref{fig:explicit-2-loop} is a 2-loop diagram.} which is achieved by the substitution $k \rightarrow k - l$ of the $k$ loop momentum integral.\footnote{Note that translations in the loop momentum $l$ are immaterial. Indeed, our analysis is to determine whether the momentum of the collinear-to-soft gluon line is pinched in the Glauber region, while such translations merely correspond to the relabelling of the momentum of that gluon line.} This will be demonstrated in this section, and we will further show that for the crossed, i.e. $q \leftrightarrow q'$, version of the diagram in \FIG\ref{fig:explicit-2-loop} (which is nothing else than topology (f) of 
\FIG\ref{fig:reduced-diagrams}), the Glauber pinch can be avoided by routing $l$ in the ``correct'' way. This is consistent with the claim made in \SEC\ref{sec:graph-f} that there is no (non-spurious) Glauber pinches corresponding to the reduced diagram in \FIG\ref{fig:reduced-diagrams} (f).

First, consider \FIG\ref{fig:explicit-2-loop}, and take the soft scaling for the $k$ and $l$ momenta, i.e.~$k,\,l \sim (\lambda,\lambda,\lambda)$, as was done in \SEC\ref{sec:pinch-analysis}. There, we also showed that  the $l^-$ component is pinched to be of ${\cal O}(\lambda^2)$ from the propagators coming from the collinear-to-$A$ subgraph, a result that we will use here. Then, in order to show that $l$ is pinched in Glauber region, it remains to argue that $l^+$ is also trapped to be $O(\lambda)$.

A simplified version of this diagram is shown in the top diagram of \FIG\ref{fig:momentum-rerouting-diag-e}, where the hard sub-process is shrunk to a point. The green arrows on the lines show the routing of the $l$ momentum, while the arrows next to the lines show the direction of flow of the $n$-momentum component (or equivalently the $-$ component). The purple arrows correspond to large components, which are fixed from the external kinematics (recall $q^-, q'^- > 0$, and $q^-, q'^- \sim \lambda^0$), while the red [blue] arrows correspond to the soft momentum component, which is of $O (\lambda)$, for the case $k^-<0$ [$k^->0$].

If the routing of the momentum $l$ is such that it goes both along and against the flow of $-$ momentum, then there are poles on opposite sides of the contour in $l^+$. To see this, consider a generic propagator with momentum $r + l$, with $|r^-| \gg |l^-|$. Writing
\begin{align}
    (r + l)^2 + i\epsilon = 2r^{-}(l^+ + ... + \sgn(r^-)i\epsilon )\,,
\end{align}
we find that the position of the $l^+$ pole in the complex plane depends on the sign of $r^-$. If $r^- > 0$, which means that the routing of $l$ momentum is along the flow of the $n$-momentum component, then the $l^+$ pole lies below the real axis. On the other hand, if $r^- <0$, which means that the routing of $l$ momentum is opposite to the flow of the $n$-momentum component, then the $l^+$ pole lies above the real axis.

Therefore, in order to have poles on opposite sides of the $l^+$ contour, one needs to have at least two propagators such that the flow of the $n$-momentum component is along one and against that routing of $l$.  Thus, determining whether the $l^+$ component is pinched or not reduces to a graphical exercise.

For example, consider the top diagram of \FIG\ref{fig:momentum-rerouting-diag-e} for the standard routing for $l$ (by standard, we mean that it corresponds to the routing for $l$ in \FIG\ref{fig:explicit-2-loop}). In that case, the two relevant propagators (i.e. those that have poles in $l^+$) are
\begin{align}
\! (l + k -q)^2 \!+ i\epsilon  &= 2(k -q)^- [l^+ \! + O(\lambda) +  \sgn(k -q)^- i\epsilon]\,, \\
      (k+l)^2 + i\epsilon  &= 2k^- [l^+ + O(\lambda) +  \sgn(k)^- i\epsilon]\,.
      \label{eq:k-plus-l-pole}
\end{align}
Since $(k -q)^- < 0$, the routing of $l$ momentum is against the flow of the $n$-momentum component in the $(l + k -q)$ propagator. Hence, the $l^+$ pole of this propagator is above the real axis. Similarly, for \EQ\eqref{eq:k-plus-l-pole}, for $k^->0$,  the routing of the $l$ momentum is along the flow of the $n$-momentum component in the $(k+l)$ propagator. Hence, the $l^+$ pole of this propagator is below the real axis.

To summarize, the crucial criterion for the existence of a Glauber pinch is that there are no possible routings of the $l$ momentum such that the flow is \textit{always} along or against the flow of the $n$-momentum component. From \FIG\ref{fig:momentum-rerouting-diag-e}, one finds that
\begin{itemize}
    \item for $k^->0$ (blue), both routings of $l$ momentum produce poles on opposite sides of the contour, which implies that the $l^+$ component is pinched. This is the Glauber pinch we identify in this paper.
    \item for $k^-<0$ (red), there exists a routing (through the meson system) of the $l$ momentum such that the poles in $l^+$ lie on the same side of the contour. This corresponds to any of the possible routings shown in the two diagrams in \FIG\ref{fig:momentum-rerouting-diag-e}.
\end{itemize}

To contrast with the above pinched configuration, we now turn to the case where the incoming photon is swapped with the outgoing photon (which corresponds to diagram (f) in \FIG\ref{fig:reduced-diagrams}). The possible routings of $l$ momentum in this case are shown in \FIG\ref{fig:momentum-rerouting-diag-f}. In this case,
\begin{itemize}
    \item for $k^- > 0$ (blue), there exists a routing (through photon vertex) for which the flow of $l$ is always along the flow of $n$-momentum. This is shown in the top diagram in \FIG\ref{fig:momentum-rerouting-diag-f}. Note that if the other routing for $l$ in the bottom diagram is chosen, one would find a pinch in $l^+$, which is of course spurious. 
    \item for $k^- < 0 $ (red), there exists a routing (through meson system) for which the flow of $l$ is always along the flow of $n$-momentum. This is shown in the bottom diagram in \FIG\ref{fig:momentum-rerouting-diag-f}. Again, if the other routing for $l$ in the top diagram is chosen, one would find a spurious pinch in $l^+$.
\end{itemize}

Therefore, we find that the configuration corresponding to diagram (f) in \FIG\ref{fig:reduced-diagrams} does not have a Glauber pinch. This is consistent with the claim made in \SEC\ref{sec:reduceddias}. Indeed, there cannot be a Glauber pinch for diagram (f) since it has only 1 incoming external particle connected to a collinear subgraph.

\begin{figure}
    \centering
    \includegraphics[width=8.5cm]{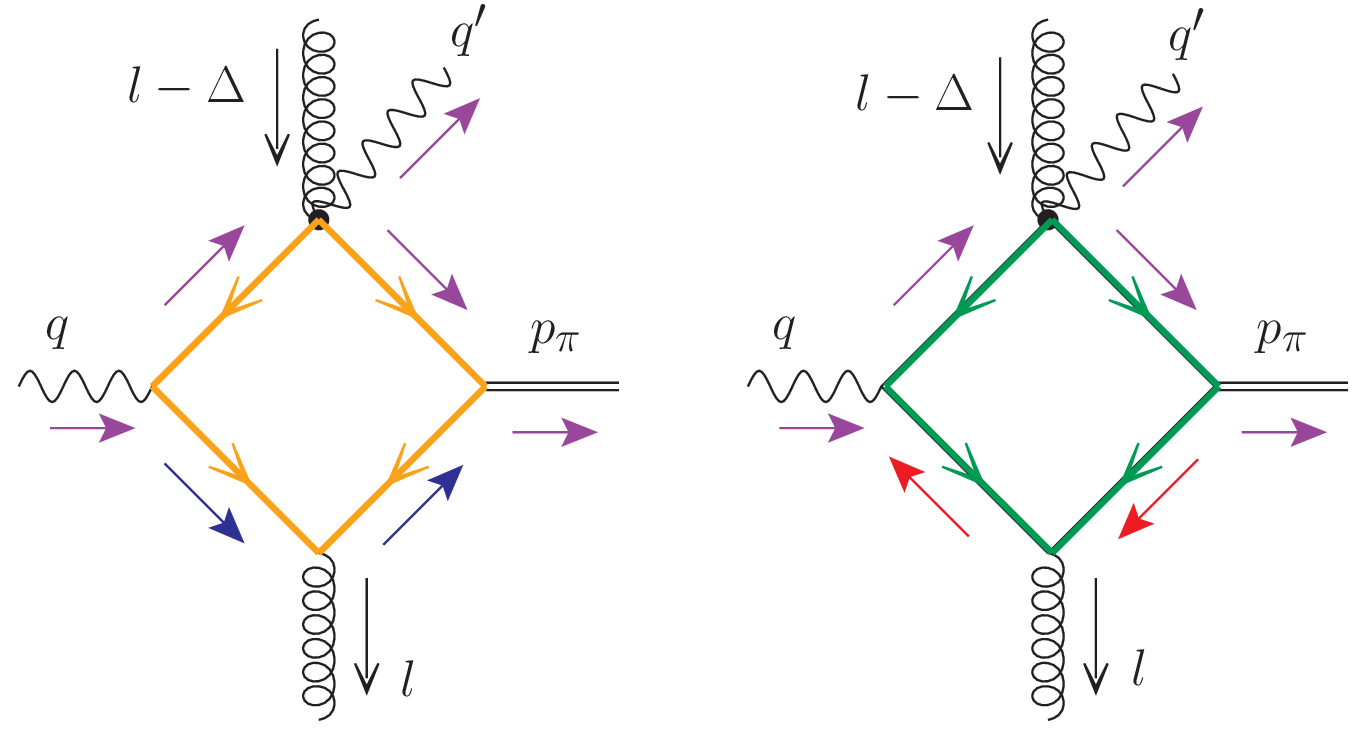}
    \\[15pt]
     \includegraphics[width=8.5cm]{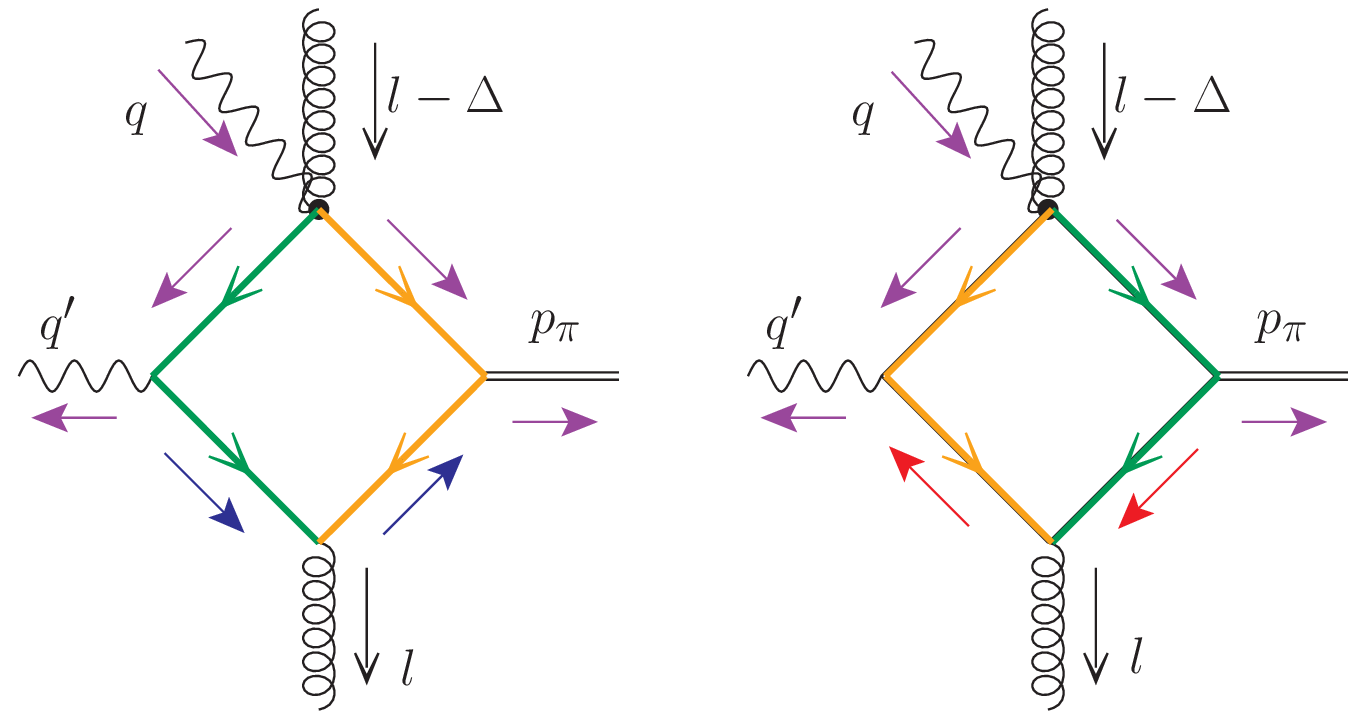}
    \caption{Simplified versions of diagram (e) [top] and diagram (f) [bottom]  used to investigate the possible routings of the $l$ momentum. The left and right panels correspond to $k^->0$ and $k^-<0$ respectively. As in \FIG\ref{fig:momentum-rerouting-diag-e}, the purple arrows show the flow of large $\sim \lambda^0$ $n$-momentum component, while the blue (red) arrows indicate the flow of small $\sim \lambda^1$ $n$-momentum component for $k^->0$ ($k^-<0$). The flow of the $l^+$ loop momentum component is from top to bottom. Routes that do not have a pinch are colored in green while routes that do are colored in orange.
    Except for the case $k^->0$ in diagram (e) (top left diagram), there is no pinch in $l^+$, since a routing for $l$ exists such that the flow of $l$ is always along or always against the flow of the $n$-momentum component. }
    \label{fig:rerouting-simplified}
\end{figure}

To illustrate the points made in this section further, we find it useful to draw the diagrams for (e) and (f) in the simplified form in \FIG\ref{fig:rerouting-simplified}. The top and bottom panels correspond to diagrams (e) and (f) respectively. The cases with $k^->0$ (blue) and $k^-<0$ (red) are considered separately, with the flow of $n$-momentum component shown in purple (large component), and red/blue (small component). In each case, there are two ways to route the flow of $l$ momentum, either by going left or right from the top. There is a pinch in $l^+$ if any routing of the $l$ goes both against and along the flow of the $n$-momentum component. It is clear from \FIG\ref{fig:rerouting-simplified} that $l^+$ is pinched only in the top left panel (i.e. the diagram corresponding to \FIG\ref{fig:reduced-diagrams} (e) with $k^- > 0$). For all other cases, a routing for $l$ exists such that $l^+$ is not pinched. The unpinched routings are shown in green, while the pinched ones are shown in orange.
	
	\subsection{Power counting for the collinear region}
\label{sec:pc-coll-region}

	Let us first derive the canonical scaling in $  \lambda  $ in the case of the collinear pinch. Such a case corresponds to $ k \sim ( \lambda ^2, 1,  \lambda ) $ and $ l \sim (1, \lambda ^2, \lambda ) $. We also project onto the \textit{transverse} polarizations of the gluons, i.e. we pick the indices $  \mu, \mu ', \nu, \nu ' $ to be transverse, by virtue of the WIs. This then implies\footnote{We separate the factors according to the following scheme
\begin{align*}
F \sim \text{momentum space volume} \times  \frac{\text{numerator}}{\text{denominator}}.
\end{align*}}
	\begin{align}
 \label{eq:pc-collinear-scaling}
F_{A}^{ \mu'_{\perp} \nu'_{\perp}} &\sim  \lambda^4 \frac{ \lambda ^2}{ \lambda ^6}= \lambda ^0\,,\qquad F_{B} \sim  \lambda ^4 \frac{ \lambda ^3}{ \lambda ^6}= \lambda ^1\,,\nonumber\\[5pt]
F_{H}^{ \mu_{\perp}  \nu_{\perp} } &\sim  \lambda ^{0}\frac{\lambda ^{0}}{\lambda ^{0}} = \lambda^0\;,
	\end{align}
giving an overall scaling of $\lambda$. This exactly corresponds to the leading power scaling we found in \SEC\ref{sec:reduceddias} for graphs (a) and (b).

	\subsection{Power counting for the Glauber region} 
	\label{sec:pc-glauber}
	Next, we consider the scalings\footnote{The same conclusion in this section would be reached for any on-shell scaling of the type $k \sim \lambda \left(\lambda^a, 1, \lambda^{\frac{a}{2}} \right) $, where $0\leq a \leq 1$. They are all equivalent to the same singularity, in the sense that they all correspond to the same expansion of the two-loop amplitude analyzed in \EQ\eqref{eq:2-loop-amplitude}. Note that for $a > 0$, this scaling corresponds to a soft-collinear scaling.}
 \begin{align}
     k \sim  \left(  \lambda,  \lambda , \lambda \right), \qquad l \sim  \left(  \lambda , \lambda ^2, \lambda  \right).
     \label{eq:Glauber-scaling}
 \end{align}
 We have argued in \SEC\ref{sec:reduceddias} that the $K$ term for both gluons vanishes by the Ward identity. Therefore, we may take the $\mu$ and $\mu'$ indices to be transverse in \FIG\ref{fig:explicit-2-loop}. We stress, however, that the same cannot be done for the Glauber gluon carrying momentum $l$. Indeed, in \SEC\ref{sec:WIs}, we will see that, for the Glauber scaling, the component of the metric tensor contracting $\nu$ and $\nu'$ that gives the leading contribution (after the sum over graphs) is
 \begin{align}
G_{\nu \nu'} = - \frac{n_{\nu'} l_{\perp \nu}}{l^+} + ... \, ,
 \end{align}
 where $...$ denotes terms that do not contribute to the leading power. To summarize, to get the leading contribution, we should count with $\mu,\mu',\nu = \perp$ and $\nu' = +$, for which we get
		\begin{align}
		 \label{eq:pc-glauber-scaling}
		F_{A}^{ \mu'_{\perp} +} &\sim  \lambda^4 \frac{ \lambda ^1}{ \lambda ^6}= \lambda ^{-1}\,,\quad
		F_{B} \sim  \lambda ^3 \frac{ \lambda ^3}{ \lambda ^4}= \lambda ^2\,,\nonumber\\[5pt]
		F_{H}^{ \mu_{\perp} \nu_{\perp} } &\sim  \lambda ^{2} \frac{ \lambda }{ \lambda^{3} }= \lambda ^0\;.
	\end{align}
	Hence, we find that the overall power of this contribution is $\lambda$, which is the leading power. It is worthwhile to point out that, when the index $  \nu' $ is transverse, the scaling for $F_{A}^{ \mu'_{\perp} \nu'_{\perp}} \sim  \lambda ^{0} $, which implies suppression in that particular case. The crucial point is that with the Glauber scaling, one can take the index $\nu'=+$, which actually leads to an enhancement by one power of $\lambda$ in the collinear-to-A sector of the amplitude, even though $F_B$ is suppressed by one power of $\lambda$  (c.f.~\EQs\eqref{eq:pc-collinear-scaling} and \eqref{eq:pc-glauber-scaling}). Therefore, the amplitude as a whole gets the same leading power of $\lambda$ as the collinear region in \SEC\ref{sec:pc-coll-region}. 
	
	Finally, we note that such an analysis takes into account the soft-end suppression from the pion distribution amplitude, since $ \phi_{\pi} \sim \frac{1}{f_{\pi}} F_{B} \sim  k^{-} \sim  \lambda$ (where we used $f_{\pi} \sim \lambda Q$). This is discussed in more detail in \SEC\ref{sec:soft-end-suppression}.

  \subsection{Power counting for the soft region}
  
  \label{sec:power-counting-soft-region}

To illustrate our argument at the beginning of \SEC\ref{sec:graph-e} that the soft region gives a power-suppressed contribution, we show here explicitly that when
\begin{align}
k \sim (\lambda_s, \lambda_s, \lambda_s), \qquad l \sim (\lambda_s, \lambda_s, \lambda_s),
\end{align}
where
\begin{align}
\label{eq:pc-soft-scaling}
\lambda \gtrsim \lambda_s \gtrsim \lambda^2,    
\end{align}
the diagram in \FIG\ref{fig:explicit-2-loop} scales as $O(\lambda^2)$. This again relies on the WIs, and the related discussion around \EQ\eqref{eq:SA-K-term}, which implies that for the sake of power counting, we can take the $\mu, \mu', \nu, \nu'$ indices to be transverse. This fact holds irrespective of the exact size of $\lambda_s$ satisfying \EQ\eqref{eq:pc-soft-scaling}. We thus obtain
 	\begin{align}
  \label{eq:soft-scaling-power-counting}
 F_{A}^{ \mu'_{\perp} \nu'_{\perp}} &\sim  \lambda_s^3 \frac{ \lambda_s^1}{ \lambda_s ^4}=  \lambda^0\,,\qquad F_{B} \sim  \lambda_s^3 \frac{ \lambda_s \lambda^2}{ \lambda_s ^4}= \lambda ^2\,,\nonumber\\[5pt]
 F_{H}^{ \mu_{\perp}  \nu_{\perp} } &\sim  \lambda_s^2\frac{\lambda_s}{\lambda_s ^3} = \lambda^0\;,
 	\end{align}
which gives a total power of $\lambda^2$. We remark that one can also take different $\lambda_s, \lambda_s'$ for $k$ and $l$ respectively. It is easy to show that this gives $ \lambda^2 \frac{\lambda_s \lambda_s'}{\max(\lambda_s^2, \lambda_s'^2)}$, such that the overall scaling is much smaller than $\lambda^2$ unless $\lambda_s \sim \lambda_s'$.

 \section{Ward identities in the Glauber region}
 \label{sec:WIs}

 \begin{figure}
		\vspace{0.2cm}
		\centering	{\includegraphics[width=6cm]{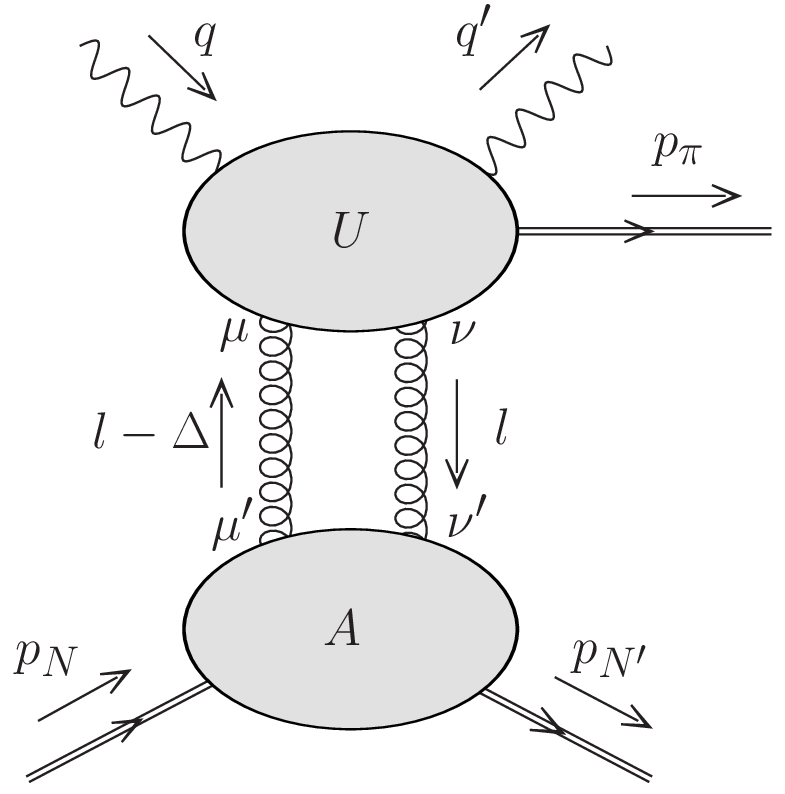}}
		\caption{Class of graphs relevant for the WI cancellations of \FIG\ref{fig:reduced-diagrams}(e). In this context, we can identify $U = H \times B \times C \times S$.}
		\label{fig:UA}
		\end{figure}

We illustrate explicitly how the WIs fail to give an extra suppression when the gluon has Glauber scaling.  For this, we have to extend the analysis beyond the single graph in \FIG\ref{fig:explicit-2-loop}, which has the Glauber pinch, to all graphs that contribute at the same order in $\alpha_s$. We stress that this is an essential step for the argument presented in this work, since even though we could get a leading power for a particular graph, as was done in \SEC\ref{sec:pc-glauber}, this does not imply that the sum over graphs would also have the same leading power. Non-trivial cancellations between graphs arise due to gauge invariance, which manifests itself through WIs. We will argue that the leading power contribution of the Glauber region persists after the sum over graphs. 

To start, consider \FIG\ref{fig:UA}, and let $F_U^{\mu \nu}$ be the contribution from the subgraph $U$ and $F_A^{\mu \nu}$ be the contribution from the subgraph $A$, which includes the gluon propagators, as well as the corresponding phase space measures $dl^-\,d^2l_{\perp}$ as in \EQ\eqref{eq:FA}. An example of subgraphs $U$ and $A$ corresponds to the diagram in \FIG\ref{fig:explicit-2-loop}, where $U = H \times B \times C \times S$. Therefore,  we can identify $F_A$ with the expression in \EQ\eqref{eq:FA} and $F_U^{\mu \nu} = \tr  \left[  {F} _{B}  {F}_{H}^{ \mu  \nu }  \right]$, with $F_B$ and $F_H^{ \mu  \nu }$ being defined in \EQ\eqref{eq:FB} and \EQ\eqref{eq:FH} respectively.

 Now, 	in the Feynman gauge, the amplitude reads
	\begin{align}
		  \sum_{A,U} \int \, F_{ A,  \mu \nu } F_{U}^{ \mu  \nu } =\int \widehat A_{\mu \nu} \widehat U^{\mu \nu} \;,
	\end{align}
 where the sum $\sum_{A,U}$ goes over the set of subgraphs $A$ and $U$ containing the case of \FIG\ref{fig:explicit-2-loop} and also, relative to this diagram, all possible connections of the two $A$-to-$U$ gluons into the upper and lower subgraphs. Moreover, we have defined $\widehat A^{\mu \nu} = \sum_A F_A^{\mu \nu}$ and $\widehat U^{\mu \nu} = \sum_U F_U^{\mu \nu}$. 
 The Ward identities are
 \begin{align}
     0 &= l_{\nu} \widehat{A}^{ \mu\nu} = l'_{\mu} \widehat{A}^{\mu \nu}  \notag
     \\
     &= l_{\nu} \widehat{U}^{ \mu\nu} = l'_{\mu} \widehat{U}^{\mu \nu},
     \label{eq:WIs}
 \end{align}
 where we defined $l' = l - \Delta$. \EQ\eqref{eq:WIs} can be readily checked by explicit calculation (without performing any integration, since it holds at the level of the integrand, provided we can translate the loop momenta internal to $A$ and $U$ respectively). It implies that the $K$ term vanishes in the sum over graphs, which is in line with the statement that at least two gluons connecting the subgraphs must be $G$-gluons, see \SEC\ref{sec:reduceddias}. Note that for more than two gluons, any additional $K$-gluon contribution would not be zero, but it would instead give the same result as the emission of $K$-gluons from a Wilson line in the adjoint representation.

Using  \EQs\eqref{eq:FA}, \eqref{eq:FB} and \eqref{eq:FH}, the superficial power counting analysis done in \SEC\ref{sec:pc-glauber} for the Glauber scaling in \EQ\eqref{eq:Glauber-scaling} gives\footnote{The subtlety that $F_H^{\mu \nu}$ has some additional suppression for $\nu = +$ for the particular graph in \FIG\ref{fig:explicit-2-loop} is irrelevant for the following argument. It will not be so for the other graphs, so that $\widehat U^{\mu\nu}$ will have the scaling in \EQ\eqref{eq:superficial-pc}. }
 \begin{align}
     \widehat{U}^{\mu \nu} \sim \lambda^2, \qquad \widehat{A}^{\mu \nu} \sim \lambda^{-2} p^{\mu} p^{\nu},
     \label{eq:superficial-pc}
 \end{align}
 where $p \sim (1, \lambda^2, \lambda)$ is a generic $\bar n$-collinear momentum and it is implied that all components of $\widehat{U}^{\mu \nu}$ scale as $\lambda^2$.

 However, by virtue of \EQ\eqref{eq:WIs} these estimations might be smaller due to cancellations in the sum over graphs. For instance, since $l'$ is $\bar n$-collinear, we get
 \begin{align}
 \label{eq:U-WI-first-index}
     \widehat{U}^{- \nu}  &= - \frac{l'_{\perp i}}{l'^+} \widehat{U}^{i \nu}  -  \frac{l'^-}{l'^+}  \widehat{U}^{ + \nu} \sim \lambda^3\,,  \\
    \label{eq:A-WI-first-index}
   \widehat{A}^{+ \nu }  &= - \frac{l_{\perp j} }{l^-} \widehat{A}^{j \nu} -  \frac{l^+ }{l^-}  \widehat{A}^{ - \nu} \sim \lambda^{-2}p^{\nu}\,,
 \end{align}
i.e.~the $\mu = -$ component in  $\widehat{U}^{\mu \nu}$ is suppressed compared to the superficial diagram-by-diagram estimate in \EQ\eqref{eq:superficial-pc}. This result justifies taking the gluon carrying momentum $l'$ in \FIG\ref{fig:explicit-2-loop} as being transverse, as was done in \SEC\ref{sec:example}.

The crucial point now is  that for the Glauber scaling in \EQ\eqref{eq:Glauber-scaling}, the WIs in \EQ\eqref{eq:WIs} do not give additional suppression. Indeed, we have
\begin{align}
\label{eq:U-WI-second-index}
    \widehat{U}^{\mu - }  &= - \frac{l_{\perp j}}{l^+} \widehat{U}^{\mu j}  -  \frac{l^-}{l^+}  \widehat{U}^{ \mu+}  \sim \lambda^{2+\delta^{\mu-}},
    \\
    \label{eq:A-WI-second-index}
    \widehat{A}^{\mu + }  &= - \frac{l_{\perp j} }{l^-} \widehat{A}^{\mu j } -  \frac{l^+ }{l^-}  \widehat{A}^{ \mu -} \sim \lambda^{-2}p^{\mu}\,.
\end{align}
The $\delta^{\mu -}$ that appears in the power of $\lambda$ in \EQ\eqref{eq:U-WI-second-index} is just a consequence of the WI in \EQ\eqref{eq:U-WI-first-index}. One therefore concludes that taking the index $\nu = -$ in $\widehat{U}^{\mu \nu}$ gives no suppression, in contrast with \EQ\eqref{eq:U-WI-first-index}, where $\mu = - $ in $\widehat{U}^{\mu \nu}$ resulted in suppression by one power of $\lambda$.

Note that when $\mu$ is transverse, the superficial power counting estimate in \EQ\eqref{eq:superficial-pc} is unchanged for the Glauber scaling of $l$.
Therefore, \EQs\eqref{eq:U-WI-second-index} and \eqref{eq:A-WI-second-index} imply that $\int \widehat U^{i-} \widehat A^{i+} \sim \lambda^1$, which is the leading power. 

In other words, the combination $\widehat A_{\mu \nu} \widehat U^{\mu \nu}$ only suffers from a single power of $\lambda$ of suppression after using the WIs (due to the $l'$ gluon momentum), such that its overall scaling is then the leading power $\lambda$, like the graphs (a) and (b) in \FIG\ref{fig:reduced-diagrams}.

To illustrate this further, we note that, as stated earlier, the $K$ term cancels in the Grammer-Yennie decomposition in \EQ\eqref{eq:GY-AH} by virtue of the WIs. We are thus left with the $G$ tensor, defined in \EQ\eqref{eq:GY-decomp-1}, which can be written as
\begin{align}
G^{\alpha \beta}(l) = g_{\perp}^{\alpha \beta} + \bar n^{\alpha} n^{\beta} - n^{\alpha} n^{\beta} \frac{l^-}{l^+} - \frac{n^{\alpha} l_{\perp}^{\beta}}{l^+}.
\label{eq:G decomp}
\end{align}
Using \EQs\eqref{eq:superficial-pc} and \eqref{eq:G decomp}, it can be readily shown that
\begin{align}
\label{eq:glauber-scaling-contractions}
&\widehat A^{\mu' \nu'} G_{\mu' \mu}(l') G_{\nu' \nu}(l) \widehat U^{\mu\nu} \notag
\\
&= - \widehat A^{\mu' \nu'} \Big ( g_{\perp\mu' i} - \frac{n_{\mu'} l'_{\perp i}}{l'^+} \Big ) \frac{n_{\nu'} l_{\perp j}}{l^+} \widehat U^{i j} + O(\lambda^2).
\end{align}
Note that the contribution from the upper subgraph $ U$ can be written entirely in terms of its transverse components. This corresponds to the fact that one can calculate the hard scattering coefficient with transverse gluon external states. \EQ\eqref{eq:glauber-scaling-contractions} clearly shows that the Glauber contribution scale as $\lambda$. We highlight that this is due to the $\frac{n_{\nu'} l_{\perp j}}{l^+}$ term, which is not suppressed when considering the Glauber scaling $l \sim (\lambda, \lambda^2,\lambda)$, compared to the collinear scaling situation, where the $\frac{n_{\mu'} l'_{\perp i}}{l'^+} $ term is suppressed by one power of $\lambda$, since $l' \sim (1,\lambda^2,\lambda)$. This justifies taking the indices in \SEC\ref{sec:pc-glauber} to be $\mu,\,\nu,\,\mu' = \perp$ and $\nu' = +$ to perform the power counting for the Glauber region.

\section{Perturbative analysis of endpoint behavior}
\label{sec:soft-end-suppression}

The parton distributions like DA ($\phi_{\pi}$) and GPDs ($F_q, F_g$) usually have a well-established endpoint behaviour when the corresponding light-cone component of a parton vanishes. For example, $\phi_{\pi}(z) \sim z$ as $z \rightarrow 0$ and $F_q(x,\xi)$ and $F_g(x,\xi)$ are non-zero and continuous, but non-analytic, at $x = \pm \xi$ \cite{Diehl:2003ny, Belitsky:2005qn}. These properties can be obtained, e.g. through studying the scale evolution. In this section, we perform simple perturbative analyses and show that they are consistent with the above properties that are supposedly valid beyond perturbation theory.

\subsection{DA}
Consider first the pion DA, which we represent by the expression of $F_B$ in \EQ\eqref{eq:FB} integrated over $k^+$ and $k_{\perp}$,
\begin{align}
\phi(k^-) &= f_{\pi}^{-1}\int \! dk^+ d^2 k_{\perp}\nonumber \\
& \!\!\!\!\times  \frac{ \slashed{k} \slashed{\cal B} \left(  \slashed{p}_{ \pi }- \slashed{k}   \right)   }{(k^2 + i \epsilon)  \left( (p_{ \pi }-k)^2 + i \epsilon \right) \left((p_B-k)^2 + i \epsilon \right)  },
\label{eq:FB-scaling}
\end{align}
where it should be recalled that ${\cal B} \sim (\lambda^3, \lambda, \lambda^2)$ and the measure of $d^4p_B$ is contained in ${\cal B}$. As indicated, we treat $\phi(k^-)$ as a function of $k^-$ (where $k^- = z p_{\pi}^-$) and we should therefore evaluate the integral on the RHS at a fixed $k^-$.

Firstly, using $p_{\pi}^- > p_B^- > 0$, it easy to see, from the pole structure or from the Landau analysis done in \SEC\ref{sec:bubble-diag-pinch}, that $\phi(k^-)$ is only non-zero if $p_{\pi}^- > k^- > 0$, which reproduces the well-known support properties of the DA. It is also easy to see that outside of the end-point region, i.e.~for $k^- \sim \lambda^0$, we have a singularity corresponding to the collinear pinch $k \sim (\lambda^2 , 1 , \lambda)$. Power-counting for this region gives $\phi(k^-) \sim \lambda^0$.

Let us now focus on the end-point region, and consider two cases, $k^- \sim \lambda$ and $k^- \sim \lambda^2$. We note that for any Glauber scaling for $k$, we can always perform a contour deformation such that $k$ is soft and usoft respectively. This is because the poles in $k^+$ from the $(k-p_{\pi})^2$ and the $(k-p_B)^2$ propagators lie on the same side of the $k^+$ contour when $k$ is (u)soft. To illustrate this explicitly, consider a soft scaling for $k \sim (\lambda, \lambda,\lambda)$:
\begin{align}
    (k-p_{B,\pi})^2+i\epsilon =0  \implies k^+ = {{\cal O}(\lambda^2)}+i\epsilon\,.
\end{align}
The $k^2$ propagator on the other hand has a pole in $k^+$ given by
\begin{align}
    k^2 + i \epsilon = 0 &\implies k^+ = -\frac{k_{\perp}^2}{2k^-} - \sgn(k^-) i\epsilon\nonumber\\
    &\implies k^+ = {\cal O}(\lambda) - \sgn(k^-) i \epsilon\,.
\end{align}
Thus, we find that the pole in $k^+$ lies on the opposite side of the contour for $k^- >0$. However, the pole of the $k^2$ propagator is not `strong' enough to pinch $k^+$ to be much smaller than $\lambda^1$. Hence, in order to obtain the scaling of $\phi(k^-)$, we should assume the scalings $k \sim (\lambda, \lambda, \lambda)$ and $k \sim (\lambda^2, \lambda^2, \lambda^2)$ respectively.

Consider first the case where $k$ is usoft, i.e. $k \sim (\lambda^2, \lambda^2, \lambda^2)$. Na\"ively, we can pick in the numerator $\slashed k {\cal B}^- \gamma^+ \slashed p_{\pi, \perp} \sim \lambda^{4}$, which gives the estimate $\phi \sim f_{\pi}^{-1} F_B \sim \lambda^{-1} \lambda^6 \frac{\lambda^{ 4}}{\lambda^8} \sim \lambda$. This appears to contradict $\phi(k^-) \sim k^-$ as $k^- \rightarrow 0$. Indeed, one should investigate the numerator more carefully. Note that after integrating over $p_B$ (recall that the measure $d^4 p_B$ is implicitly included in $\cal B$, such that the mass dimension of $\cal B$ is two), we must have, by dimensional analysis and Lorentz covariance,
\begin{align}
\int \frac{{\cal B}^{\mu}}{(p_B-k)^2} = \alpha p_{\pi}^{\mu} + \beta  k^{\mu},
\label{eq:B-scaling-after-integration}
\end{align}
where $\alpha\sim \lambda^{-1}$ and $\beta = \mathcal O(\lambda^{-1})$, whose scalings are obtained by relating the power of $\lambda$ for each index $\mu$. Since $k$ is usoft, we can neglect the second term, so that after inserting \EQ\eqref{eq:B-scaling-after-integration} in \EQ\eqref{eq:FB-scaling}, we get
\begin{align}
\label{eq:soft-end-suppression}
\phi &\sim f_{\pi}^{-1} \int dk^+ d^2 k_{\perp}  \frac{ \slashed{k} \slashed p_{\pi}  \slashed{p}_{ \pi }  \alpha }{k^2  \left( p_{ \pi }-k \right)^2 }  \notag
\\
&\sim \lambda^{-1} \lambda^6  \frac{\lambda^4 \lambda^{-1}}{\lambda^6} \sim \lambda^2.
\end{align}
In other words, for usoft $k$, we get an additional suppression from the numerator due to having only a single external momentum available, namely the momentum of the pion. What happens is that the terms $\slashed k {\cal B}^- \gamma^+ \slashed p_{\pi, \perp}$ and $\slashed k \slashed  {\cal B}_{\perp} \gamma^+ p^-_{\pi}$ in the numerator of \EQ\eqref{eq:FB-scaling}, each scaling as $\lambda^4$, cancel each other, such that the numerator then effectively scales as $\lambda^5$.

The same also holds for the soft case, i.e.~when $k \sim (\lambda,\lambda,\lambda)$. The numerator in \EQ\eqref{eq:soft-end-suppression} changes to $\slashed{k} (\alpha\slashed p_{\pi} + \beta\slashed{k} )( \slashed{p}_{ \pi } -\slashed{k})$, with $\alpha\sim \lambda^{0}$ and $\beta = \mathcal O(\lambda^{1})$.
Repeating the same analysis as in the usoft case, we get
\begin{align}
\label{eq:alt-soft-end-suppression}
\phi &\sim f_{\pi}^{-1} \int dk^+ d^2 k_{\perp}  \frac{ \slashed{k} (\alpha\slashed p_{\pi} + \beta\slashed{k} )( \slashed{p}_{ \pi } -\slashed{k})  }{k^2  \left( p_{ \pi }-k \right)^2   }  \notag
\\
&\sim \lambda^{-1} \lambda^3  \frac{\lambda^2}{\lambda^3} \sim \lambda\,.
\end{align}
This concludes the discussion of the DA.

\subsection{Quark GPD}
We will now discuss the analogous case for the quark GPD, which will nicely illustrate the difference in endpoint behaviour compared to the DA. Let
\begin{align}\label{eq:Fqlplus}
&F_q(l^{+}) = \int dl^- d^2l_{\perp}\nonumber\\
&\times \frac{\slashed{l} \slashed{\cal A} (\slashed{p}_{N'} -\slashed{p}_N - \slashed{l}) }{(l^2+i \epsilon) ((p_{N'}-p_{N}-l)^2+i \epsilon) ((p_A-l)^2+i \epsilon)},
\end{align}
where ${\cal A} \sim (1, \lambda^2, \lambda)$ is a collinear subgraph which contains the measure of the $p_A \sim (1, \lambda^2, \lambda)$ loop momentum. In \SEC\ref{sec:coll pinch} and \APP\ref{app:collinear-pinch-Landau}, we discuss the pinching of the $l^-$ contour with the important conclusion, that in addition to the collinear, soft and ultrasoft pinch we also have a Glauber pinch $l \sim (\lambda, \lambda^2, \lambda)$ if  we fix\footnote{A similar pinch $l \sim (\lambda^2, \lambda^2, \lambda)$ would appear if we fix $l^+ \sim \lambda^2$.} $l^+ \sim \lambda$. Thus, we consider all of these four cases in the following.

As in the DA case, the numerator ${\cal A}$ in \EQ\eqref{eq:Fqlplus} contains the measure $d^4p_A$. Repeating the integral performed in \EQ\eqref{eq:B-scaling-after-integration} for the quark GPD case at hand, we get
\begin{align}
\int \frac{{\cal A}^{\mu}}{(p_A-l)^2} = \alpha p_{N}^{\mu} + \alpha' p_{N'}^{\mu}+ \beta  l^{\mu}\,.
\label{eq:A-scaling-after-integration}
\end{align}
Let us start with the collinear scaling, i.e. $l \sim (1,\lambda^2,\lambda)$. By equating the power counting on either side of \EQ\eqref{eq:A-scaling-after-integration} for different $\mu$, we obtain $\alpha,\,\alpha',\,\beta \sim \lambda^{-2}$. Plugging this into \EQ\eqref{eq:Fqlplus} gives
\begin{align}
    F^{\mathrm{coll.}}_q(l^+) &\sim \int dl^- d^2l_{\perp} \nonumber\\
    &\quad\frac{\slashed{l} (\alpha \slashed{p}_N + \alpha' \slashed{p}_{N'} + \beta \slashed{l}) (\slashed{p}_{N'} -\slashed{p}_N - \slashed{l}) }{l^2 (p_{N'}-p_{N}-l)^2}\nonumber\\
    &\sim \lambda^4 \frac{ \lambda^{-2}\lambda^2}{\lambda^4} \sim \lambda^0\,.
\end{align}
As expected, the quark GPD scales as $\lambda^0$ when $l$ has collinear scaling.

Consider now the soft scaling for $l \sim (\lambda,\lambda,\lambda)$. \EQ\eqref{eq:A-scaling-after-integration} remains the same, but this time, $\alpha, \alpha' \sim \lambda^{-1}$ and $\beta \sim {\cal O}(\lambda^{-1})$. Then, the scaling for $F_{q}(l^+)$ becomes
\begin{align}
    F^{\mathrm{soft}}_{q}(l^+) \sim \lambda^3 \frac{\lambda^{-1} \lambda^2}{\lambda^3} \sim \lambda^1\,.
\end{align}
Similarly, the ultrasoft scaling for $l \sim (\lambda^2,\lambda^2,\lambda^2)$ also gives the decomposition in \EQ\eqref{eq:A-scaling-after-integration}, with $\alpha,\,\alpha' \sim \lambda^{-2}$ and $\beta \sim {\cal O}(\lambda^{-2})$. We then obtain
\begin{align}
    F^{\mathrm{usoft}}_{q}(l^+) \sim \lambda^6 \frac{\lambda^{-2} \lambda^3}{\lambda^6} \sim \lambda^1\,.
\end{align}
For both soft and usoft scalings, the result for the scaling of $F_{q}(l^+)$ is somewhat unexpected, since we would na\"ively expect the quark GPD to be non-vanishing when $l$ becomes (ultra)soft.

Finally, we take the $\bar{n}$-coll.-to-soft Glauber scaling $l \sim (\lambda,\lambda^2,\lambda)$. Now, $\alpha,\alpha' \sim \lambda^{-2}$ and $\beta \sim {\cal O}(\lambda^{-2})$, which leads to 
\begin{align}
    F_{q}^{\mathrm{Glau.}}(l^+) \sim \lambda^{4} \frac{\lambda^{-2}\lambda^{2}}{\lambda^{4}} \sim \lambda^0\,.
\end{align}
Thus, we find that, unlike the soft and usoft scalings, the Glauber scaling for $l$ actually gives a contribution to $F_{q}(l^+)$ of order $\lambda^0$. Therefore, we conclude that, from a perturbative point of view, the fact that the quark GPD is non-vanishing at the breakpoints $x \pm \xi$ is fundamentally connected to the Glauber region. 

We highlight that in the DA case, Glauber regions can all be deformed to the soft regions, which implies that the soft suppression obtained in \EQs\eqref{eq:soft-end-suppression} and \eqref{eq:alt-soft-end-suppression} is indeed correct.

	\section{Conclusions}
	\label{sec:conclusions}

In this work, we present a detailed analysis of the exclusive photoproduction of a $\pi^0\gamma$ pair with large invariant mass. We identify all the leading power contributions, and show that, in addition to the standard collinear contributions (quark and gluon channels), there exists a contribution of the same power which involves a Glauber gluon exchange between the collinear sector defined by the incoming and outgoing nucleons, and the soft sector connecting the incoming photon and outgoing meson. We show explicitly that this Glauber contribution is pinched, implying that it cannot be deformed to other regions. The Glauber pinch that we have identified relies on  two different loop momenta, which are intertwined in a specific way that makes the pinch appear. Through a careful study of a generic 2-loop example shown in \FIG\ref{fig:explicit-2-loop}, we verify that the power counting of such a contribution matches the collinear one, and is therefore leading. Moreover, for this analysis, we also considered all possible 2-loop diagrams connected to that in \FIG\ref{fig:explicit-2-loop} by different gluon attachments, in order to verify that the identified Glauber region does not become suppressed due to Ward identity cancellations.

Furthermore, our results imply that the corresponding crossed process of $\pi^0$-nucleon scattering to two photons also suffers from the same issue, since the Glauber pinch is also present there. One important and peculiar aspect of our findings is that the pinched Glauber gluon in our analysis corresponds to one of the two active gluons that would usually participate in the hard partonic scattering level -- In fact, it is precisely for  this reason that the amplitude diverges if one attempts to compute this contribution, already at leading order and leading twist, when collinear factorization is na\"ively assumed.

However, we stress that for cases where the gluon exchange channel is forbidden, either due to electric charge conservation in the case of charged meson production, or due to C-parity conservation in the case of neutral vector meson production, the Glauber pinch does not exist at leading power, and collinear factorisation at leading twist is expected to work without issues. To save the phenomenology for the photoproduction of a $\pi^0\gamma$ pair, it is necessary to go \textit{beyond} standard collinear factorization - The natural approach is to introduce $k_T$-dependent distributions. We intend to address this issue in the future.

Finally, we remark that there has been progress in a new approach for determining pinch singularities through tropical geometry, see e.g. \cite{Pak:2010pt,Jantzen:2012mw} and for some recent developments \cite{Gardi:2022khw,Fevola:2023kaw,Fevola:2023fzn}. While this approach is very promising, because it potentially gives a systematic way to find all regions, there is still some substantial development required. It would, however, be interesting to address the issues encountered in this work using this formalism.
 
	\section*{Acknowledgements}
	
	We would like to thank Renaud Boussarie, Volodya Braun, John Collins, Markus Diehl, Goran Duplan\v{c}i\'{c}, Alfred Mueller, Melih Ozcelik, Kornelija Passek-Kumeri\v{c}ki, Bernard Pire, George Sterman, Iain Stewart, Jianwei Qiu, Zhite Yu  for many useful discussions.
 This work was supported by the GLUODYNAMICS project funded by the ``P2IO LabEx (ANR-10-LABEX-0038)'' in the framework ``Investissements d’Avenir'' (ANR-11-IDEX-0003-01) managed by the Agence Nationale de la Recherche (ANR), France. This work was also supported in part by the European Union’s Horizon 2020 research and innovation program under Grant Agreements No. 824093 (Strong2020). This work was partly supported by the Science and Technologies Facilities Council (STFC) under grant  ST/X00077X/1. This project has also received funding from the French Agence Nationale de la Recherche (ANR) via the grant ANR-20-CE31-0015 (``PrecisOnium'')  and was also partly supported by the French CNRS via the COPIN-IN2P3 bilateral agreement. J.S. was supported in part by the Research Unit FOR2926 under grant 409651613, by the U.S. Department of Energy through Contract No. DE-SC0012704 and by Laboratory Directed Research and Development (LDRD) funds from Brookhaven Science Associates.
 J.S. also acknowledges the hospitality of IJCLab where part of this work was done. The work of L. S. is supported by the grant 2019/33/B/ST2/02588 of the National Science Center in Poland. L. S. thanks the P2IO Laboratory of Excellence (Programme Investissements d’Avenir ANR-10-LABEX-0038), the P2I - Graduate School of Physics of Paris-Saclay University, and IJCLab for support.

 \appendix

 \section{Illustration of the breakdown of collinear factorization at leading order}
\label{app:calc}
\begin{figure}[th!]
		\vspace{0.2cm}
		\centering	{\includegraphics[width=7cm]{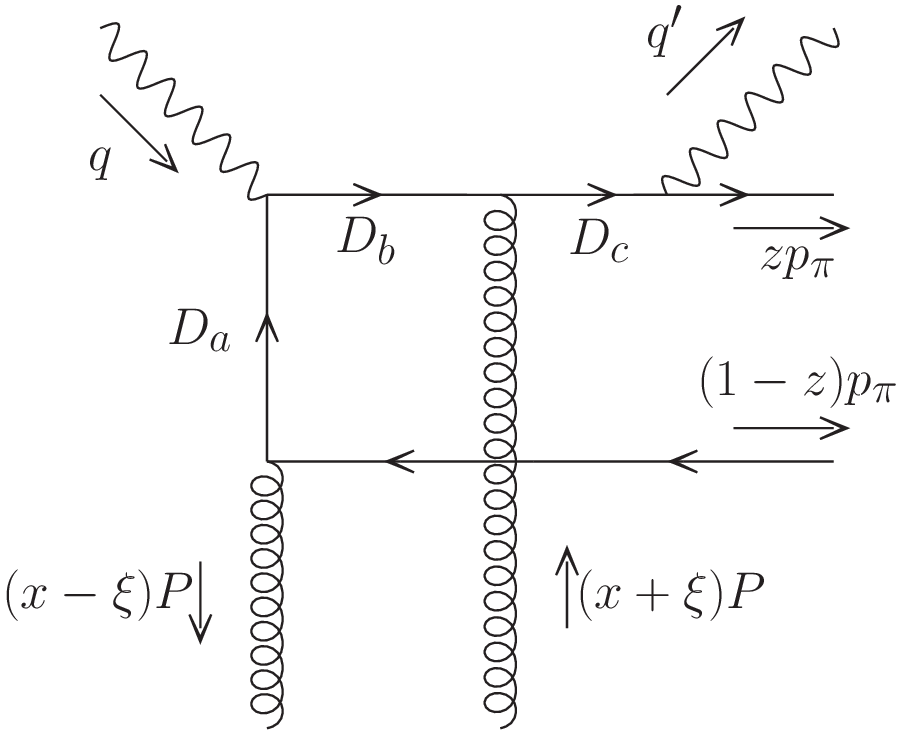}}
		\caption{A particular Feynman diagram contributing to the coefficient function that causes a divergence upon the double convolution over the momentum fractions $x$ and $z$ with the GPD and DA.}
		\label{fig:diag-10}
		\end{figure}

Let us assume collinear factorisation na\"ively, and calculate the gluon-induced amplitude for  photoproduction of a $  \pi ^{0}\gamma  $ pair. This calculation results in divergences in the double integration over the momentum fractions in specific Feynman diagrams which exactly correspond to the topologies where the Glauber pinch occurs, i.e.~\FIG\ref{fig:reduced-diagrams}(e). For example, the diagram in \FIG\ref{fig:diag-10}, which corresponds to \FIG\ref{fig:reduced-diagrams}(e) in the limit $x \to \xi$ and $\bar{z} \to 0$, where $\bar{z} \equiv 1-z$, is given by
	\begin{widetext}
			\begin{align}
			\hspace{-2cm} {\cal M } &\propto	  	
			\int_{-1}^{1} dx \int _{0}^{1}dz
			 \frac{\mathrm{Tr} \left[  \slashed{p}_{\pi}\gamma ^{5} \slashed{\epsilon }_{q'} \left(  \slashed{q}'+z  \slashed{p}_{\pi}   \right)\gamma ^{j} \left(  \slashed{q}-(x-\xi) \slashed{P}-\bar{z} \slashed{p}_{\pi}    \right) \slashed{\epsilon }_{q} \left( -(x-\xi) \slashed{P}-\bar{z} \slashed{p}_{\pi}   \right) \gamma ^{i}      \right] }{ \left[ 2z\,q' \cdot   p_{\pi}  \right]  \left[ -2 \left( x-\xi \right)q  \cdot  P - 2 \bar{z}\,q  \cdot  p_{\pi}+2\bar{z} \left( x-\xi \right)P   \cdot p_{\pi}+i\epsilon    \right]   \left[ 2\bar{z} \left( x-\xi \right)P  \cdot  p_{\pi}+i\epsilon   \right] }\nonumber\\[5pt]
			&\qquad\times   \frac{\phi_{\pi} (z) H_{g}(x, \xi ) g_{ij}^{\perp}}{ \left[ x- \xi +i\epsilon  \right]  \left[ x+ \xi -i\epsilon  \right] } \nonumber \\[7pt]
			&\hspace{-0.5cm}\stackrel{x\to \xi,\bar{z}\to 0}{\longrightarrow}\;\propto
			\int_{-1}^{1} dx \int _{0}^{1}dz\frac{(x-\xi) \bar z}{ \left[  \left( x-\xi \right)+ A \bar{z} - i\epsilon   \right] \left[(x-\xi) \bar z + i\epsilon\right]\left[  x-\xi +i\epsilon   \right]   }\,,\qquad A >0.
   \label{eq:explicit-result}
		\end{align}
	\end{widetext}
		In the above expression, we have neglected overall prefactors, and we have introduced the positive constant $ A \sim  {\cal O }(1)  $ that corresponds to a ratio of scalar products of momenta of on-shell external particles, noting that all such scalar products are positive. We have also focused on the projection onto the unpolarized gluon GPD $ H_{g} $ to reduce clutter in the above expression (the same conclusion can be reached with the polarized gluon GPD $ \tilde{H}_g $ as well). The two denominators that accompany the gluon GPD $H_g(x,\xi,t)$ in \EQ\eqref{eq:explicit-result} come from relating the matrix element of two gauge fields to the gluon GPD $H_g$, which is defined from the matrix element of two gluon field strengths,
\begin{align}
&\int_{-\infty}^{\infty} \frac{dz^-}{2\pi} e^{ixz^- P^+} \langle p_{N'} | A_a^{i}(-z^-) A_b^{j}(z^-) | p_{N} \rangle \notag
\\
&= -\frac{\bar{u}(p_{N'})\gamma^{+} u (p_{N})}{4P^+}\frac{\delta_{ab}\, g_{\perp}^{ij} }{N_c^2 - 1} \frac{ H_g(x,\xi, t) }{(x-\xi\pm i\epsilon)(x + \xi \mp i\epsilon)}  + ...
\label{eq:AA in terms of GPD}
\end{align}
The choice of the sign of $i\epsilon$ corresponds to the residual gauge ambiguity of expressing the nucleon matrix element of two gauge fields in terms of the universal GPD $H_g$, which is gauge invariant since it is defined from two gluon field strengths (see \APPs D and G in \cite{Belitsky:2005qn}). The conventional choice, e.g. in DVCS, is $[x-\xi+i\epsilon][x+\xi-i\epsilon]$.
However, neither sign will prevent the divergence from occurring.\footnote{Even if $[x-\xi-i\epsilon]$ is chosen, the pinch will persist. This can be seen by writing the numerator as two parts, $[(x-\xi) \bar z+i\epsilon]-[i\epsilon]$. The first term will of course be finite, but the second term, while having a factor of $i\epsilon$ in the numerator, is actually divergent logarithmically as $\log(\epsilon)$ upon performing the double integration. This can be understood by the fact that partial fraction gives rise to $1/(i\epsilon)$ terms in this case.} Finally, we highlight that the signs of all other $i\epsilon$ factors in \EQ\eqref{eq:explicit-result} are fixed from the Feynman prescription.

In the second line of \EQ\eqref{eq:explicit-result}, we took the limit $ x \to  \xi $ (soft gluon) and $\bar{z} \to 0  $ (soft anti-quark), which corresponds to the region in \FIG\ref{fig:reduced-diagrams}(e). Such a region is of course unavoidable from the integration over the momentum fractions. Finally, for simplicity, we have fixed the functional form of the distribution amplitude to be $  \phi_{\pi} (z) \sim z  \bar{z} $.
	
	Therefore, one finds that the amplitude for the diagram in \FIG\ref{fig:diag-10} has a divergent imaginary part,
	\begin{align}
		\label{eq:divergence-diag-10} 
&\int_{-1}^{1} dx \int _{0}^{1}dz\frac{x-\xi }{ \left[  \left( x-\xi \right)+ A \bar{z} - i\epsilon   \right] \left[x-\xi + i\epsilon\right]\left[  x-\xi + i \epsilon   \right]   }\nonumber
\\[5pt]
&\qquad \propto\, i\log(\epsilon) + O(\epsilon^0)\;.
	\end{align}
	The pinching of the two poles in the above equation comes from the two quark propagators $D_a$, which becomes \textit{soft}, and $D_b$, which becomes \textit{collinear} to the incoming photon, in the limit  $x \to \xi$ and $\bar{z} \to 0$.
 
	Moreover, the \textit{full} amplitude can be computed, and it can be verified explicitly that no cancellations happen among the complete set of 24 Feynman diagrams that contribute to this process. This is of course to be expected by the general arguments made in this work, see also \SEC\ref{sec:WIs}.
 The full amplitude, corresponding to the projection onto the gluon GPD $ H_g $, is given by
\begin{widetext}

\begin{align}
	\label{eq:divergence-full-amplitude}
	\hspace{-0.8cm}\sum_{ \mathrm{all} }  {\cal M } &\propto \frac{   f_{\pi}\left( x^2-\xi^2 \right) \left[ - \alpha  \left[  \left( x^2-\xi^2 \right)^2  \left( 1-2z \bar{z} \right)+8 x^2  \xi ^2 z \bar{z}   \right]- \left( 1+ \alpha ^2 \right)z \bar{z}  \left( x^4-\xi^4 \right)    \right]}{z\bar{z} \left[ x- \xi +i\epsilon  \right] \left[ \bar{z} \left( x+ \xi  \right)- \alpha z \left( x- \xi  \right) -i\epsilon   \right]\left[ z \left( x- \xi  \right)+ \alpha \bar{z} \left( x+ \xi  \right) -i\epsilon   \right]  }\nonumber\\[7pt]
	&\times \frac{1}{  \left[ x+ \xi -i\epsilon  \right] \left[ \bar{z} \left( x- \xi  \right)+ \alpha z \left( x+ \xi  \right) -i\epsilon   \right]\left[ z \left( x+ \xi  \right)- \alpha \bar{z} \left( x- \xi  \right) -i\epsilon   \right]}\times \frac{H_{g}(x,\xi) \, \phi_{\pi} (z)}{\left[ x- \xi +i\epsilon  \right]\left[ x+ \xi -i\epsilon  \right]}\nonumber\\[9pt]
	&	\hspace{-0.5cm}\stackrel{x\to  \xi, \bar{z}\to 0}{\longrightarrow}\propto\frac{ - \alpha  \left[   \left( x-\xi \right)^2 +
2 \xi ^2  \bar{z}
		\right]- \xi \left( 1+ \alpha ^2 \right) \bar{z}  \left( x-\xi \right)   }{ \left[ x- \xi +i\epsilon  \right] \left[  \left( x- \xi  \right)- 2\frac{\xi}{\alpha} \bar{z} +i\epsilon   \right]\left[ \left( x- \xi  \right)+ 2\xi \alpha \bar{z} -i\epsilon   \right]  }\;.
\end{align}
\end{widetext}
	In the above, we have again taken the functional form of the distribution amplitude to be $  \phi_{\pi} (z) \sim z \bar{z} $ in the second line, and introduced the dimensionless parameter 
 \begin{align}
     \alpha \equiv \frac{(p_{\pi}-q)^2}{(p_{\pi}+q')^2}\,,
 \end{align}
 which is neither very small nor very close to 1. In any case, the divergence will be independent of the exact value of $\alpha$. We note that the full amplitude is symmetric under $ x \to -x $ and $ z \to \bar{z} $. Thus, if the divergence persists at $ x \to \xi,\, \bar{z} \to 0 $, it will also be present in three other regions corresponding to applyinh the previously-mentioned symmetries.
	
In this way, we find that the divergence illustrated in \EQ\eqref{eq:divergence-diag-10} is still present in \EQ\eqref{eq:divergence-full-amplitude}, due to the last term in the square brackets in the numerator, $ 2 \xi ^2 \bar{z} $, which has only a single power of zero in the limit $ x \to \xi,\, \bar{z} \to 0 $. This necessarily causes two propagators to be `pinched', as in \EQ\eqref{eq:divergence-diag-10}.

It is worthwhile to point out that at the level of the full amplitude in \EQ\eqref{eq:divergence-full-amplitude}, the denominators that accompany the gluon GPD definition in \EQ\eqref{eq:AA in terms of GPD} cancel out (there is an explicit factor of $x^2 - \xi^2$ in the numerator in the first line of \EQ\eqref{eq:divergence-full-amplitude}). This cancellation is expected due to Ward identities, i.e. gauge invariance, see the discussion in \SEC\ref{sec:WIs}.
Therefore, the divergence that one observes in the full amplitude upon double integration over $x$ and $z$ is completely independent of the way that the $\frac{1}{x - \xi}$ and $\frac{1}{x+\xi}$ factors are regulated\footnote{However, the choice of the sign of $i\epsilon$ matters at the level of individual diagrams.} in the definition of the gluon GPD, see the text below \EQ\eqref{eq:AA in terms of GPD}.

Finally, we remark that it should be possible to get a well-defined asymptotic expansion of the diagram in \FIG\ref{fig:explicit-2-loop}, i.e.~the integral in \EQ\eqref{eq:2-loop-amplitude} (in the collinear limit, its hard part reduces to \FIG\ref{fig:diag-10}). Rigorously, this can be done in dimensional regularization, e.g. using the method of regions formalism. It might be possible that other regulators are needed to give the contribution from each region a well-defined expression and possibly one also has to take into account the overlap contributions.  By taking all the contributions into account, the divergence in \EQ\eqref{eq:divergence-full-amplitude} would not appear. Therefore, the appearance of the divergence in the amplitude assuming collinear factorisation is a direct consequence of the existence of other regions (which we have identified as a Glauber region here) to the leading term in the power expansion in $\lambda$ of the full amplitude. Such an analysis is beyond the scope of this paper.

\section{A simple 1D example illustrating a pinch and power counting}

\label{app:simple-1D-example}

\begin{figure}
     \centering
     \includegraphics[scale=.5  ]{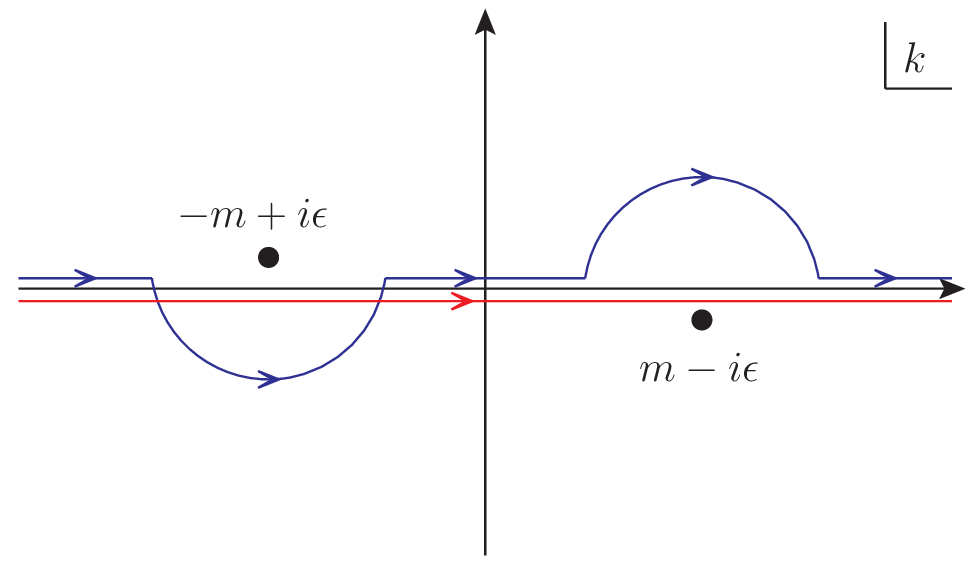}
     \caption{The integral $I_0(m)$ in the complex plane. The poles at $k = \pm m$ approach the real axis as $\epsilon \rightarrow 0^+$. We can deform the contour from the real axis (red) to the shown toy contour with half circles centered at $k = \pm m$ (blue).}
     \label{fig:poles}
 \end{figure}

 As a trivial example, consider the integral
 \begin{align}
 \label{eq:I0-m}
   I_0(m) &= \lim_{\epsilon \rightarrow 0^+} \int_{-\infty}^{\infty} dk \, \frac{1}{(k-m+i\epsilon)(-k-m+i\epsilon)} \notag
\\
    &= \frac{i\pi}{m}.
 \end{align}
For $m \neq 0$, both denominators cannot be zero at the same time. Therefore, the Landau condition for a single denominator can clearly not be fulfilled.

On the other hand, for $m = 0$, the Landau condition reads 
\begin{equation}
\label{eq:Landau-conditions-I0-m}
\alpha_1 - \alpha_2=0, \quad \alpha_1,\alpha_2 \geq 0,     \quad \alpha_1+\alpha_2 >0\,,
\end{equation}
where $\alpha_1$ corresponds to the $(k-m+i \epsilon)$ denominator, and $\alpha_2$ to the $(-k-m+i \epsilon)$ denominator. \EQ\eqref{eq:Landau-conditions-I0-m} has  solutions $\alpha_1=\alpha_2 > 0$, implying the existence of a pinch at $k = 0$. This corresponds of course to the pole of $I_0$ at $m = 0$ in \EQ\eqref{eq:I0-m}. 

For one-dimensional integrals such as $I_0$, it is easy to visualize the pinch by considering the contour in the complex plane, see \FIG\ref{fig:poles}. The poles coalesce as $m \rightarrow 0$ and $\epsilon \rightarrow 0^+$ so that the integration contour cannot be deformed away.

While in this trivial case, it is easy to evaluate the integral by closing the contour at infinity in either the upper or lower half plane and using the residue theorem, this is not so for general classes of Feynman diagrams.
In the following, we will illustrate the concept of \textit{power-counting analysis} which, when applied to $I_0$, can be used to determine the $\frac{1}{m}$ behaviour generically without evaluating the integral.

The argument is as follows. The integration contour is forced to pass in between the two poles, with a \textit{maximum} distance of $m$ from both poles. It is thus clear that one needs to investigate the contribution to the integral in a neighborhood of this approximate pinch. To estimate the value of the integral, we investigate an integration region with a  size $\sim m$ around the poles. To this end, we need to find a correct estimator for the integrand inside that integration region. This is achieved by choosing the blue contour in \FIG\ref{fig:poles}, since the integrand has the same order of magnitude $\sim \frac{1}{[m][m]}$ everywhere in that integration region in that case. The estimate for the integral $I_0$ then becomes
\begin{align}
    I_0(m) \sim \frac{[m]}{[m][m]} \sim \frac{1}{m}\,,
\end{align}
which matches the scaling in \EQ\eqref{eq:I0-m} as expected. 

Note that what we have done is basically dimensional analysis. To see this, rescale $k \rightarrow mk$, to get an overall $\frac{1}{m}$ multiplying an $m$-independent integral, which is finite by the Landau condition, and which has to be non-zero, since the contour passes in between two poles.

\section{Collinear pinch for the off-forward kinematics using the Landau condition}
\label{app:collinear-pinch-Landau}

We re-derive the result in \EQ\eqref{eq:cond1} by using the Landau equations. First consider the general case of having two on-shell propagators $(k-q_1)^2 = 0$ and $(k-q_2)^2 = 0$, where $q_1^2 = q_2^2 = q_1 \cdot q_2= 0$. Then, it is easy to see that the Landau equations reduce to
\begin{align}
k  = \alpha q_1 + (1-\alpha) q_2,
\end{align}
where $1 \geq \alpha \geq 0$. Without loss of generality, take $q_1$ and $q_2$ to be  collinear in the $\bar n$ direction. Then $k^+ = \alpha q_1^+ + (1-\alpha) q_2^+$ and therefore the Landau equations have a solution if and only if 
\begin{align}
\max (q_1^+, q_2^+ ) \geq k^+ \geq \min(q_1^+, q_2^+). 
\end{align}
Note that the equalities coincide with soft singularities of one of the propagators.
Applying this result to each pair of propagators in \EQ\eqref{eq:l denoms} immediately gives \EQ\eqref{eq:cond1} as a necessary and sufficient condition for the collinear pinch.

	\bibliographystyle{utphys}
	
	\bibliography{masterrefs.bib}

\end{document}